\documentclass[preprint]{aastex}

\tighten
\input psfig.tex

\newcommand{\asec}{\hbox to 1pt{}\rlap{$^{\prime\prime}$}.\hbox to 2pt{}}
\newcommand{\amin}{\hbox to 1pt{}\rlap{$^{\prime}$}.\hbox to 2pt{}}

\slugcomment{Submitted to {\it The Astronomical Journal}}
\shortauthors{Lauer et al.}
\shorttitle{The Centers of Early-Type Galaxies}

\begin{document}

\title{The Centers of Early-Type Galaxies with {\it HST.}
V. New WFPC2 Photometry
\footnote{Based on observations made with the NASA/ESA 
{\it Hubble Space Telescope}, obtained at the Space Telescope Science Institute,
which is operated by the Association of Universities for
Research in Astronomy, Inc., under NASA contract NAS 5-26555. These
observations are associated with GO and GTO proposals
\# 5236, 5446, 5454, 5512, 5943, 5990, 5999, 6099, 6386, 6554, 6587, 6633,
7468, 7820, 8683, and 9107.}}

\author{Tod R. Lauer}
\affil{National Optical Astronomy Observatory\footnote{The National Optical
Astronomy Observatory is operated by AURA, Inc., under cooperative agreement
with the National Science Foundation.},       
P.O. Box 26732, Tucson, AZ 85726}

\author{S. M. Faber}
\affil{UCO/Lick Observatory, Board of Studies in Astronomy and
Astrophysics, University of California, Santa Cruz, CA 95064}

\author{Karl Gebhardt}
\affil{Department of Astronomy, University of Texas, Austin, TX 78712}

\author{Douglas Richstone}
\affil{Department of Astronomy, University of Michigan, Ann Arbor, MI 48109}

\author{Scott Tremaine}
\affil{Princeton University Observatory, Peyton Hall, Princeton, NJ 08544}

\author{Edward A. Ajhar}
\affil{Department of Natural Sciences, Mathematics, and Computer Sciences,
St. Thomas University, Miami Gardens, FL 33054}

\author{M. C. Aller}
\affil{Department of Astronomy, University of Michigan, Ann Arbor, MI 48109}

\author{Ralf Bender}
\affil{Universit\"{a}ts-Sternwarte, Scheinerstra\ss e 1, M\"{u}nchen 81679,
Germany}

\author{Alan Dressler}
\affil{The Observatories of the Carnegie Institution of Washington,
Pasadena, CA 91101}

\author{Alexei V. Filippenko}
\affil{Department of Astronomy, University of California, Berkeley,
CA 94720-3411}

\author{Richard Green}
\affil{National Optical Astronomy Observatory, P.O. Box 26732, Tucson, AZ 85726}

\author{Carl J. Grillmair}
\affil{SIRTF Science Center, Pasadena, CA 91125}

\author{Luis C. Ho}
\affil{The Observatories of the Carnegie Institution of Washington,
Pasadena, CA 91101}

\author{John Kormendy}
\affil{Department of Astronomy, University of Texas, Austin, TX 78712}

\author{John Magorrian}
\affil{Department of Physics, University of Durham, Durham, United Kingdom,
DH1 3LE}

\author{Jason Pinkney}
\affil{Department of Physics and Astronomy, Ohio Northern University,
Ada, OH 45810}

\author{Christos Siopis}
\affil{Department of Astronomy, University of Michigan, Ann Arbor, MI 48109}

\vfill

%\altaffiltext{2}{The National Optical Astronomy Observatory is
%operated by AURA, Inc., under cooperative agreement with the National
%Science Foundation.}       

\begin{abstract}

We present observations of 77 early-type galaxies imaged with the PC1 CCD
of {\it HST}+WFPC2.  ``Nuker law'' parametric fits to the surface brightness
profiles are used to classify the central structure into ``core'' or
``power-law'' forms.  Core galaxies are typically rounder than power-law
galaxies.  Nearly all power-laws with central ellipticity $\epsilon\geq0.3$
have stellar disks, implying that disks are present in most power-laws with
$\epsilon<0.3,$ but are not visible due to unfavorable geometry.  A few
low-luminosity flattened core galaxies also have disks; these may be transition
forms from power-laws to more luminous core galaxies, which lack disks.  Several
core galaxies have strong isophote twists interior to their break radii,
although power-laws have interior twists of similar physical
significance when the photometric perturbations implied by the twists are
evaluated.  Central color gradients are typically
consistent with the envelope gradients; core galaxies have somewhat weaker
color gradients than power-laws.  Nuclei are found in 29\%\ of the cores
and 60\%\ of the power-laws.  Nuclei are typically bluer than
the surrounding galaxy.  While some nuclei are associated with AGN, just as many
are not; conversely, not all galaxies known to have low-level AGN exhibit
detectable nuclei in the broad-band filters.  NGC 4073 and 4382,
are found to have central minima in their intrinsic starlight distributions;
NGC 4382 resembles the double nucleus of M31.  In general, the peak
brightness location is coincident with the photocenter
of the core to a typical physical scale $<1$ pc. Five
galaxies, however, have centers significantly displaced from their surrounding
cores; these may be unresolved asymmetric double nuclei.  Lastly, as noted by
previous authors, central dust is visible in about half of the galaxies.  The
presence and strength of dust correlates with nuclear emission, thus
dust may outline gas that is falling into the central black hole.  The
prevalence of dust and its morphology suggest that dust clouds form,
settle to the center, and disappear repeatedly on $\sim10^8$ yr timescales.  We
discuss the hypothesis that cores are created by the decay of a massive black
hole binary formed in a merger.  Apart from their brightness profiles,
there are no strong differences between cores and power-laws that
demand this scenario; however, the rounder shapes of cores, their lack of disks,
and their reduced color gradients may be consistent with it.

\end{abstract}

\keywords{galaxies: nuclei --- galaxies: photometry --- galaxies: structure}

\section{Introduction} 

This paper presents {\it HST/WFPC2} observations of 77 early-type galaxies.
We investigate the central morphology of the galaxies by characterizing
the properties of their nuclei, color gradients, ellipticities,
isophote twists, instances of central surface brightness minima,
offset centers, dust content, and dust morphologies.  The result is a more
complete portrait of the sample galaxies than has been available heretofore and
a reference where many different properties are measured and compared in
one place.  We also derive high resolution surface photometry profiles
of the sample galaxies, which we characterize with ``Nuker law'' fits
\citep{l95}.  Analysis of the fit parameters is presented in Paper VI
(Lauer et al., in preparation), in which we combine the present results
with previously published Nuker law fits to {\it HST} surface photometry
profiles to define an extended sample of 264 E/S0 galaxies.

A motivation of both this paper and Paper VI is to understand the origin
of ``cores'' in early-type galaxies.  {\it HST} images show that nearly
all galaxies have singular starlight distributions in the sense that
surface brightness diverges as $\Sigma(r)\sim r^{-\gamma},$ with $\gamma>0$
\citep{l91, l92a, l92b, crane, k94, f94, l95}.  In typically lower-luminosity
early-type galaxies, $\gamma$ decreases only slowly as the center is approached
and a steep $\gamma>0.5$ cusp continues into the {\it HST} resolution limit;
\citet{l95} classified these systems as ``power-law'' galaxies.  In more
luminous galaxies, however, the projected profile transitions or breaks from a
steep power law in the envelope to a shallow inner cusp with $\gamma<0.3$;
these ``core galaxies'' thus show central deficits of starlight
or a core compared to the centrally-steeper ``power-law'' galaxies.
Many of the core galaxies are the same systems in which cores
were already evident from ground-based observations \citep{k85, l85};
however, rather than representing regions in which the central stellar
density becomes constant, the residual shallow cusps in projected
brightness still imply steep and singular cusps in density \citep{l95}.

\citet{g96} and \citet{f97} showed that the distribution of cusp
slopes at the {\it HST} resolution limit in both stellar luminosity density
and projected surface brightness is bimodal; power-laws and core
galaxies are separated into two distinct groups by their inner cusp slopes.
\citet{rest} and \cite{rav} later identified a small number of
``intermediate'' galaxies that have limiting cusp slopes
with $0.3<\gamma<0.5,$ but showed that the ensemble of cusp
slopes in all early-type galaxies is still bimodal.
This topic will be further explored in Paper VI, which strongly ratifies
the bimodality of central structure.

\citet{f97} also examined how the central structure
correlates with other galaxy properties,
showing that luminous early-type galaxies preferentially have cores,
whereas most fainter spheroids have power-law profiles.
Moreover, cores are slowly rotating and have boxy isophotes,
while power laws rotate rapidly and are disky.
These ideas resonate well with a revision of
the Hubble sequence proposed by \citet{kb96}, which divided ellipticals into
boxy, non-rotating types and disky, rotating types.  The latter
were seen to be the close relatives of spirals, whereas the boxy,
non-rotating ellipticals were somehow ``different."
Cores are associated with the boxy, non-rotating sequence,
and thus serve as a fundamental morphological marker.

\citet{f97} further argued that the prevalence of cores is directly
tied to the presence of nuclear black holes in nearly all galaxies \citep{mag}
and the assembly of galaxies by hierarchical mergers.
\citet{bbr} argued that the merging of two galaxies, each harboring
a central massive black hole, would create a black hole binary
in the center of the merger remnant.
Gradual hardening of the binary via stellar encounters would scatter stars
out of the center, creating a central deficit of stars
with respect to inward extrapolation of the envelope
in a merged galaxy whose center would otherwise be very dense.
N-body simulations of merging galaxies with central black holes
\citep{ebi, mak97, qh97, mnm} show that cores can indeed form in such merger
remnants.  Looking for ways to test the hypothesis that cores reflect
the effects of binary black holes on the central structure of galaxies
is a subtext for much of the analysis presented in this paper.
We will return to this topic in the summary.

A final section of the paper is devoted to the morphology and prevalence
of optical dust absorption in early-type galaxies.  As in previous works (e.g.,
\citealt{dekoff}, \citealt{vdf}, \citealt{t01}), dust is found in roughly half
of our galaxies.  Dust is extremely well correlated with both the
presence and strength of nuclear optical
emission, which we assume indicates AGN activity.
Thus, it is hard to avoid concluding that dust is outlining
interstellar material that is about to fall onto the black hole,
and that it therefore becomes a valuable clue as to how that happens.
We discuss the possibility that the diverse patterns of dust absorption
seen in early-type galaxies may be viewed as various stages of a
``settling sequence;'' under this picture dust may come and go on
timescales of a few $\times10^8$ years.
The implication is that galaxies might be emptying and refilling
themselves with dust constantly, raising the question of where the dust
comes from and why the process is cyclical.  Understanding this could be an
important clue to the growth of black holes.

\section{Observations}

\subsection{The Sample \label{sec:samp}}

We present images and surface photometry for 77 early-type galaxies
observed with {\it HST}+ WFPC2; 55 of these were obtained under programs
GO 5512, 6099, 6587, and 9107 which were carried out by our collaboration.
The observations are listed in Table \ref{tab:obs}.
We also include an independent reduction of the galaxies observed
in GO 5454 (PI: Franx), which were selected to have
kinematically decoupled cores (KDC).  Analysis of these
by \citet{carollo} showed no morphological differences between ordinary
ellipticals and those with KDC, thus including these galaxies in
the present sample causes no obvious bias.
Some of the remaining galaxies represent original or archival images
obtained in support of GO 9107, in which we proposed STIS spectroscopy to
explore the mass range of nuclear black holes at a particular galaxy
luminosity.  Lastly, there are additional galaxies obtained
from the archive in support of our other STIS programs.
A portion of this photometry has been published previously in
papers on the detection of nuclear black holes \citep{k96,g3379,bower,g03}
or on the morphology of galaxies with double centers or local minima in
surface brightness at their centers \citep{l96,l02}; such galaxies are
presented again here to provide complete surface photometry in the
context of a comprehensive re-investigation of the central structure of
early-type galaxies.  Table \ref{tab:obs} lists any earlier publication
of surface photometry derived from the given dataset.

There is no single criterion that characterizes our sample.  In general,
the sample comprises relatively luminous nearby elliptical and
early-S0 galaxies that could be investigated spectroscopically with
{\it HST} to detect and ``weigh" nuclear black holes.  The particular
selection of galaxies was heavily informed by the central-structure
parameter relationships presented by \citet{f97}. Several galaxies were
selected, for example, within the luminosity range over which power-law
and core galaxies co-exist, to explore the transition between the two types.
Others were selected to extend the luminosity range of the core
parameter relationships. The properties of the galaxies, such as luminosity,
velocity dispersion, distance, and morphology are listed in Table
\ref{tab:glob}. A set of galaxies already observed with WFPC1
and presented by \citet{l95} were also re-observed with WFPC2 when it was
felt that the improved resolution and dynamic range might lead to a
significant improvement in the galaxies' central structure.
In the end, the sample is rich in core galaxies, with relatively few
power-law galaxies; it thus is complementary to the \citet{rest}
sample, which is rich in power-law galaxies.

\subsection{Images and Image Reduction \label{sec:red}} 

All galaxies were imaged with the high-resolution PC1 CCD in WFPC2;
the image scale is $0\asec0456/$pixel.  Nearly all galaxies were imaged
in the F555W filter (equivalent to broad band $V$);
images were also obtained in F814W filter (broad band $I$) to provide
color information for many galaxies.  The exposure
time for most galaxies was roughly a single orbit in each filter, typically
split into several sub-exposures to allow for cosmic-ray event repair.
The exposure levels in the galaxy centers varied widely over the sample,
but were typically $>10^4 e^-,$ which is optimal for point-spread function
(PSF) deconvolution.

Beginning in program G0 6587, many of the image sets were dithered in
a $2\times2$ square pattern of 0.5 PC pixel substeps.  This procedure
allows for the removal of the aliasing present in any single PC image
and consequent recovery of the optimal spatial resolution provided
by the {\it HST}+WFPC2 combination.  The reduction procedure for
a given image set depended on whether the images were dithered.
Non-dithered image sets were always obtained with the same pointing and could
simply be compared to detect and repair cosmic-ray events, before being stacked.

The dithered images for a given galaxy and filter combination
were combined into a single Nyquist-sampled ``super-image'' using the
Fourier algorithm presented by \citet{l99a}.  The algorithm combines
the images in the Fourier domain and is designed to recover the Nyquist image
even if the dither positions are somewhat non-optimal.  The output of the
algorithm is a double-sampled image with 0.5 PC pixel steps.
The Fourier algorithm is theoretically justified and introduces no
blurring of the data; this is in contrast to {\it Drizzle} \citep{driz},
which cannot remove aliasing in a small dataset, and introduces
additional blurring on the scale of the original pixel sampling.
A drawback of our reduction path is that the pointing of
each image in the image-set is slightly different, complicating the
repair of cosmic-ray events.\footnote{This is not a difficulty intrinsic to
dithering. With an additional orbit's worth of exposure time, redundant
exposures could be obtained at each dither position.}  The procedure
used for cosmic ray repair was to first intercompare the images ignoring
their small offsets and then construct a Nyquist image from the repaired images.
This first estimate of the Nyquist image was used to improve the
initial detection of the cosmic-ray events, thus producing a revised
Nyquist image.  This procedure was iterated a second time to produce
the final image.

The final step was to deconvolve the PSF.  The {\it HST+}WFPC2
PSF introduces significant blurring as far as $0\asec5$ from the galaxy
centers, but deconvolution can recover the intrinsic surface brightness
profiles to a few percent accuracy for $r\geq0\asec04$ \citep{l98}.
Deconvolution was achieved by 40 iterations of the Lucy-Richardson
\citep{lucy,rich} algorithm.  The PSFs were calculated from the Tiny-Tim
package \citep{tiny}.  For the double-sampled images, the Tiny-Tim
PSFs were convolved with the pixel-response function of \citet{l99b}.

\subsection{Surface Photometry \label{sec:surf}}

Central images of the sample galaxies are shown in Figures
\ref{fig:images} and \ref{fig:obs_im}.  While the classic view
of early-type galaxies is that they are purely stellar systems with
elliptical isophotes, the majority of the galaxies depart from this picture
one way or another, having central dust patches or rings, stellar disks,
nuclear clusters, central minima, and so on.  A summary of the
morphological features seen is presented in Table
\ref{tab:morph}.  Figure \ref{fig:images} also shows the galaxy images divided
by models reconstructed from the surface photometry
(described below) to help isolate these features.
Those galaxies that had their centers so obscured by dust that surface
photometry could not be derived are shown separately in Figure \ref{fig:obs_im}.
The overall importance of dust for any galaxy was roughly quantified
by comparing the integrated light of the observations to model galaxy images
reconstructed from the surface photometry within $1''$ of the galaxy centers.
The dust absorption thus estimated is presented in Table \ref{tab:morph} in
units of hundredths of a magnitude.  These absorption values are not intended
for quantitative astrophysical interpretation, but do seem to correspond well
to subjective impressions of the strength of central dust seen in
Figure \ref{fig:images}.

Surface photometry was derived
under the assumption that the isophotes are
concentric ellipses, but with position angle and ellipticity than can
vary arbitrarily with radius.  The photometry was measured in several stages.
The first step was to mask obvious stars, dust patches, image
defects, and so on, from the images.  Initial photometry
was derived using the least-squares algorithm of \citet{l86}.
A model reconstructed from photometry was then subtracted from the images
to isolate and mask more subtle features that were missed during the
initial pass.  For galaxies with extensive dust this procedure was
often iterated several times, as the isophote parameters would
often change significantly as the mask was refined.  The final stage
was to measure the inner isophotes ($r<0\asec5$) with the
high-resolution algorithm of \citet{l85}. This algorithm uses
sinc-interpolation to sample the image, thus preserving the
spatial resolution of the images.  For this stage, image defects were
initially filled in with a model constructed from the lower-resolution
photometry, and then with a model constructed from the high-resolution
photometry.

The final surface photometry profiles are presented in Figure \ref{fig:surf}
\footnote{Tabulated versions of the photometry are presented
in the electronic version of this paper, and at
http://www.noao.edu/noao/staff/lauer/nuker.html.}.
The zeropoint calibration was obtained from \citet{holtz}.  A ``sky''
level for each galaxy was measured from the corner of the WFPC2 WF3 CCD
farthest away from the galaxy center (typically $1\amin8$ in angle),
and was subtracted from the photometry.  For the more extended galaxies,
the sky will have some galaxy-light contamination, but given that the
present study focuses on the brighter central isophotes, this will cause
only a small error.

\subsection{WFPC1 versus WFPC2 Photometry \label{sec:wfpc1}}

A number of galaxies in the sample were originally observed with WFPC1.
The new WFPC2 photometry allows a test of the WFPC1
observations, which required large deconvolution corrections.
Figure \ref{fig:decon} compares WFPC2 and WFPC1 profiles for several galaxies.
\citet{l93} argued based on simulations that deconvolution could recover
profiles from WFPC1 that were accurate to a few percent
for $r>0\asec1,$ with significant residual blurring at smaller radii.
The comparison of WFPC2 and WFPC1 photometry for M32 and M33
presented by \citet{l98} showed this conclusion to be essentially correct,
while underscoring the need to deconvolve WFPC2 photometry as well.
Figure \ref{fig:decon} ratifies this conclusion for a larger
set of galaxies.  For the most part, the WFPC2 and WFPC1 profiles
agree extremely well for $r>0\asec1.$  The exception to this
general rule appears to be galaxies with strong nuclei,
such as NGC 3115, where light from the central point-source
spilled beyond $r\approx0\asec1,$ and was poorly recovered by WFPC1.
Weak nuclei in core galaxies, such as in NGC 1399, were
also poorly captured by WFPC1.  The conclusion, overall,
is that the WFPC1 profiles can still be used to their stated
$0\asec1$ resolution limit.

\section{Nuker Law Fits}

We fitted the surface brightness profiles
with the ``Nuker law'' \citep{l95} parametric form
\begin{equation}
I(r)=2^{(\beta-\gamma)/\alpha}I_b\left({r_b\over r}\right)^{\gamma}
\left[1+\left({r\over r_b}\right)^\alpha\right]^{(\gamma-\beta)/\alpha},
\label{eqn:nuker}
\end{equation}
which describes the profiles as a ``broken power law;''
it is an extension of a broken power-law form first used to
describe the brightness profile of M32 \citep{l92b}.
Here $r$ is the semi-major axis\footnote{This
differs from \citet{byun} where $r$ represented the geometric mean
of the semi-major and semi-minor axes.}.
The outer portion of the profile is a steep power-law
with logarithmic slope, $-\beta.$  This outer power-law
``breaks'' at radius $r_b,$ transitioning to a shallower
inner power-law with slope $-\gamma.$  The smoothness or ``speed"
of transition is determined by $\alpha;$ the intensity scale
is set by $I_b,$ the surface brightness at $r_b.$
The break radius is both the point of maximum curvature of
the profile in logarithmic coordinates and the location where the local
slope is the average, $-(\beta+\gamma)/2,$ of the inner and outer slopes.
In passing, we note that we are aware of the limitations
of parametric fits to data.
However, a nonparametric analysis of the WFPC1 profiles by
\citet{g96} yielded no additional insight into the central structure
of elliptical galaxies than was provided by the parametric analysis
of \citet{l95} and \citet{f97} using the Nuker law.

We fitted the major-axis profiles using the MINUIT package \citep{jr},
which was also used by \citet{byun} to fit the WFPC1 profiles.
Unlike Byun et al., however, we allowed the algorithm to find
solutions with $\gamma<0;$ the discovery of galaxies with
central {\it minima} in surface brightness \citep{l02} shows
that central brightness profiles with positive slopes must be allowed for.
While in formally $\gamma<0$ would imply that the surface brightness
profile goes to zero as $r\rightarrow0,$ which is impossible in projection,
the integrated brightness in any aperture remains positive, and more to the
point, we mean to imply no behavior of the profile within the resolution limit.
The Nuker law parameters for all galaxies in Figure \ref{fig:images}
are listed in Table \ref{tab:cen} and plotted in Figure \ref{fig:surf}.
The radial domain fitted was adjusted on an {\it ad hoc} basis
based on the structure of a given galaxy and the residuals of
an initial Nuker law fit over the entire extent of the profile.
Typically an inner limit was specified to avoid nuclei
(see below), which the Nuker law is not intended to represent.
An outer limit was specified in some cases to avoid deviations from
the Nuker law that might have influenced the quality of the fit
to the inner region.  The fits are shown as solid lines
in Figure \ref{fig:surf}, with the extent of the line showing the
radial range of the fit.  The fit residuals are plotted at the bottom
of the brightness profile panels.

\citet{l95} showed that the brightness profiles could be grouped
into two classes.  ``Core galaxies'' show an obvious break in logarithmic
slope that marks a transition to a shallow inner cusp with $\gamma'<0.3,$
where $\gamma'$ is the slope at the {\it HST} resolution limit
($0\asec04$ for the F555W WFPC2 images).
``Power-law galaxies,'' in contrast, retain a steep slope, $\gamma'>0.5,$
into the resolution limit.  As was shown by \citet{g96} and \citet{f97},
the distribution of cusp slopes is bimodal, with no examples
of galaxies in those papers with $0.3<\gamma'<0.5.$  This classification scheme
will be re-examined in detail in Paper VI.  The
adopted profile classifications are listed in Table \ref{tab:cen}.
In some cases, as noted in Table \ref{tab:cen},
we have reclassified a number of galaxies previously observed
by ourselves or others because
we have adopted finer resolution limits or used higher
resolution imagery than the previous investigations.

\citet{rest} and \citet{rav}
identified a set of galaxies that have $0.3<\gamma'<0.5,$
which Rest et al.\ call ``intermediate'' galaxies.
Three such galaxies (NGC 821, 3585, and 7723) are
in the present sample.  As \citet{rav} showed, intermediate types
are rare, so that the distribution of cusp slopes remains
bimodal, even if the region between core and power-law galaxies
is no longer completely empty.  We also validate the concern of \citet{rest}
that the intermediate classification is highly sensitive to resolution.
As it happens, after adopting finer resolution limits
than did Rest et al., we consider {\it all} of their intermediate
galaxies to be core galaxies, while we now consider some of
their power-law galaxies to be intermediates.  We will
explore this issue in detail in Paper VI.

\section{The Central Structure of Early-Type Galaxies}

\subsection{Inner Isophote Ellipticity and Central Stellar Disks}

Figure \ref{fig:ellip} plots the average isophote ellipticity interior to $r_b,$
weighted by the integrated isophote luminosity
(analogous to equation (\ref{eqn:lw}) in $\S\ref{sec:twist}$ below),
as a function of the average exterior ellipticity (averaged
the same way but for $r>r_b$) for both core and
power-law galaxies (ellipticity values are given in Table \ref{tab:ellip}).
There is no clear
trend for either type of galaxy to become systematically flatter or rounder
as the center is approached.  What is evident in Figure \ref{fig:ellip},
however, is that the power-law galaxies on average are more
elliptical than core galaxies.  This trend is also seen in Figure
\ref{fig:ellip_mag}, which plots inner ellipticity as a function
of galaxy luminosity.  While \citet{vincent} show that elliptical
galaxies overall become rounder with increasing luminosity,
and cores become more prevalent with luminosity, core galaxies are
rounder than power-law galaxies at the same luminosity.
\citet{ryden2} looked at {\it HST} photometry of a large sample of elliptical
galaxies and reached similar conclusions.  In particular, they looked
at isophote shape as a function of the fractional distance to the effective
radius, finding that the distinction between core and power-law galaxies
was strongest at the smallest radii.

\citet{jaffe} and \citet{f94} noted that the power-law galaxies
(their Type II) in their sample were highly elliptical, while
the core galaxies (their Type I) were nearly round.
They argued that the difference between the two was {\it merely
one of projection.} In their view, all ellipticals harbored disks,
which when seen edge-on would make sharply rising cusps; the stellar density
profiles of all ellipticals would have essentially the same form.
\citet{l95} disagreed with this picture, noting the existence of round
power-law galaxies, which should not exist at all in the projection model.
Moreover, the new data in Figures \ref{fig:ellip} and \ref{fig:ellip_mag}
show that core galaxies co-exist with power-law galaxies at all ellipticities.
Although core galaxies are on average rounder than power-law galaxies,
the ellipticity distributions of the two types overlap significantly,
whereas the projection model implies a much cleaner separation.

\citet{l85} also claimed that the majority of highly elliptical
power-law galaxies showed no evidence of a central stellar disk;
however, we now think that such cases are rare.
First \citet{f97} subsequently noted that power-law galaxies have disky
{\it outer} isophotes, which firmly associates power-laws with disks.
Now, new WFPC2 imagery here shows that disks can be seen in {\it
nearly all} power-law galaxies with inner ellipticity $\epsilon\geq0.3,$
whereas {\it no} disks are seen in power-law galaxies rounder
than that (see Figure \ref{fig:ellip_mag}).
This strong dichotomy suggests that {\it all} power-law galaxies
harbor disks; their visibility is simply due to the viewing angle.
It is true that \citet{l95} found disks in
only three out of nine highly elliptical power-law galaxies;
however, some of the disks in the present sample are faint
and may have been poorly visible in WFPC1 imagery.
While, in common with \citet{l95}, we cannot detect a disk in the
power-law galaxy NGC 1426 (which has $\epsilon=0.32$), we
do find a disk in NGC 4458 (and now classify it as a core
rather than power-law galaxy).

Core galaxies generally have boxy isophotes \citep{f97}, but
we find disks in four core galaxies, all of which have high
inner $\epsilon\geq0.3.$ Three of these are rather dim,
having $M_V>-21,$  and only one (NGC 3706) of the five highly elliptical
core galaxies brighter than this has a disk.
Two of the three intermediate galaxies also have disks,
and they also lie in the same magnitude transition zone near $M_V\sim-21.$

The overall impression is that disks are frequent if not ubiquitous
in power-law galaxies, populate a fair fraction of faint cores and
intermediate galaxies in the transition zone, and then peter out amongst
bright core galaxies.  The disappearance of disks with brightness seems
to be related to the core/power-law transition itself and may give a
clue as to how it occurs, as discussed in the summary.

\subsection{Isophote Twists in the Inner Profiles \label{sec:twist}}

An examination of the isophote position angle profiles presented
in Figure \ref{fig:surf} shows that many core galaxies
have significant isophote twists close to or interior to
their break radii.  This behavior, for example, can be seen
in NGC 507, the first galaxy presented in Figure \ref{fig:surf}.
The general phenomenon is that most core galaxies have
isophote orientations that remain essentially constant over
their outer radii (within the regions imaged by WFPC2), but at some
small radius the isophotes smoothly twist away from the outer orientations.

To quantify this phenomenon, we fitted the position angle profiles
of the core galaxies with a ``hinged'' line.  In detail, a line
was first fitted over the outer portion of the profile, starting at
an inner limiting ``hinge'' radius.  Once the parameters
of the outer line were determined, a second line was
fitted interior to the hinge radius
(but still with $r>0\asec13$ to avoid nuclei)
with the constraint that the inner line had
to intersect with the outer line at the hinge radius.  The
effect is to fit the entire position angle profile with two lines
of differing slope that intersect at a ``hinge.''  The position
of the hinge was selected by finding the location that minimized
the residuals of the entire position angle profile fitted by
both line segments.  The final fits are shown as dotted lines
in the position angle plots in Figure \ref{fig:surf}.

Table \ref{tab:bend} presents the values of the inner and outer position
angle gradients, the hinge radius, and the local logarithmic slope
of the brightness profile, $\gamma',$ at that location.
Figure \ref{fig:bend} shows the inner position angle gradients
as a function of $\gamma'$ at the transition radius.  The symbols
plotted also show the inner and outer gradients schematically
as angled wedges, with the slopes of the segments to the left
and right of the vertex proportional to the inner and outer
position-angle gradients.  Both the table and figure show that
the isophote position angle varies only slowly outside the hinge radius,
but strong twists occur at small radii.
The transition in behavior occurs near $\gamma'\approx0.5,$ a point
typically within the shallow surface brightness cusp
interior to the core break-radius.

Figure \ref{fig:bend_pow} plots the same data for the power-law galaxies.
The power-law galaxies appear to have at most very small
twists in their inner isophotes.  Only two galaxies not affected by
dust have gradients that exceed $30^\circ/(\Delta\log(r)=1).$
Consideration of the true ``power'' in the twists in power law galaxies,
however, suggests that they are as significant as the large twists
seen in the inner cusps of the core galaxies.
The significance of any given isophote twist is tied to
how much the light distribution in a galaxy must have changed to
exhibit the twist.  Very little light needs to be redistributed
in a portion of a galaxy with a shallow brightness gradient
and nearly circular isophotes to cause a strong twist in angle,
while the perturbation must be large when the gradient is steep
or the isophotes are highly elliptical.

We have attempted to quantify these effects by following the general
approach of \citet{ryden}, who analyzed the shapes of dwarf ellipticals by
introducing a twist-measure that accounts for the true perturbation
in the projected light distribution of a galaxy to produce the observed twists.
The basic concept is to measure the relative amount of light required
to ``iron-out'' or untwist a twisted galaxy.
The first step is to use the brightness profiles to calculate the
luminosity-weighted average position angle of a given galaxy,
\begin{equation}
\langle\phi\rangle={1\over L(r_1,r_2)}\int_{r_1}^{r_2}\phi(r)\Sigma(r)dA,
\label{eqn:lw}
\end{equation}
over a major-axis range $r_1<r<r_2$ in which the twist is to be measured.
Here $\phi(r)$ gives the isophote position angle profile, $\Sigma(r)$
is the brightness profile, $L(r_1,r_2)=\int_{r_1}^{r_2}\Sigma(r)dA$
is the integrated galaxy luminosity over the radial range,
and the area element is
\begin{equation}
dA=2\pi q r \left(1+{1\over2}{d\ln q\over d\ln r}\right)dr,
\end{equation}
where $q=1-\epsilon,$ the ratio of isophote minor to major axes at $r.$
The ``surface-brightness twist'' is estimated as
\begin{equation}
T=\left({1\over N}\sum\left({I_{\langle\phi\rangle}(x,y)-I(x,y)\over
I(x,y)}\right)^2\right)^{1/2},
\label{eqn:twist}
\end{equation}
where $I_{\langle\phi\rangle}(x,y)$ is an image reconstructed from the
surface photometry with the isophotes all forced to have position angle
$\langle\phi\rangle,$ while $I(x,y)$ is an image reconstructed from the surface
photometry with the twists as measured\footnote{Both reconstructions
assume concentric elliptical isophotes and allow the ellipticity to
vary with radius.},
and the sum is over the $N$ pixels within $r_1<r<r_2.$
The surface-brightness twist thus is a typical relative root-mean-square
(rms) measure of the difference between the untwisted and observed galaxy.
We emphasize that our $T$ measure differs from that of \citet{ryden},
which we believe suffers from a conceptual error in its derivation.

We have calculated surface-brightness twists separately
for the radius ranges inside and outside the hinge radius
(limited to an outer radius three times larger than
the hinge-radius defining the boundary of the inner and outer gradients);
they are presented in Table \ref{tab:bend} and Figure \ref{fig:twist}.
As can be seen, both power-law and core galaxies
exhibit somewhat stronger twists in their inner regions.
The inner-twists are typically $\sim3\times$ stronger for
both types of galaxies than the outer twists.  Displayed this way,
however, we see that the significance of the inner twists in
both core and power-law galaxies is essentially the same.

\subsection{Central Color Gradients \label{sec:colgrad}}

The $V-I$ color gradients of the inner regions of the sample galaxies
are generically similar to those seen farther out in the envelopes
of early-type galaxies.  The great majority of sample galaxies become
slightly redder toward the center.  The exceptions are NGC 2300,
4382, and 7457, which have shallow gradients that become {\it bluer} with
decreasing radius.  As seen in Figure \ref{fig:surf}, the color
gradients are roughly linear with shallow slopes
in logarithmic coordinates.  Table \ref{tab:vmi} lists the color
gradients measured over $0\asec2<r<10'';$ the inner cutoff is set
to avoid nuclear sources.  Figure 4 of \citet{carollo} suggests that the
color gradients show more variability over $0\asec25<r<1\asec5$ than they
do at larger radii, but we do not see this effect.

The strengths of the color gradients superficially appear to depend on
the profile type of the galaxy, as already suggested by Figure 11 of
\citet{carollo}, and by high-resolution ground-based images of galaxy
centers \citep{mich}; however, this effect appears largely to be due to
the greater prevalence of dust in power-law galaxies.
Figure \ref{fig:grad_gam} plots the color gradients as a
function of the limiting logarithmic profile slope
at the resolution limit, $\gamma';$ the symbol type encodes
the central dust indices given in Table \ref{tab:morph}, with
open symbols corresponding to galaxies with dust absorption
greater that 0.02 mag.  Even though we attempted to exclude visible dust
patches from the profile estimation algorithms, the strongest
color gradients correlate with the largest scatter about the mean gradient
line, supporting the conclusion that the color profiles in these cases are
still affected by residual dust absorption and are not likely to reflect
true changes in the underlying stellar population.
If dusty galaxies are ignored, there still a suggestion
of a weak trend for power-law galaxies to have stronger gradients
than core galaxies; the mean $V-I$ gradients for the two types
are $-0.045\pm0.005$ mag (excluding the one power-law galaxy with a
positive gradient) and $-0.031\pm0.006$ mag, respectively.
Clearly, however, the scatter in color gradients at any $\gamma'$
is substantially larger than the small offset between the two types,
and to first order both types of galaxies have
essentially the same inner color gradients.
Lastly, Figure \ref{fig:grad_mag} shows that there is no relationship
between the strength of the gradient and galaxy luminosity,
once dusty galaxies are excluded.

We also examined the color gradients
of core galaxies to see if the gradients were smaller interior to
the break radius.  Because the envelope color gradients in core
galaxies are small to begin with, such a change in absolute terms
will be a small effect, and indeed for the majority of core galaxies
it is difficult to see any change
in the gradient associated with the core.  However, NGC 3379, 3608,
4291, 4365, 4406, and 7619 do appear to have flatter color
gradients for $r<r_b;$ NGC 4406 provides the best example, but the
significance of this effect is low.

\subsection{Nuclei}

A substantial fraction of the sample galaxies have nuclei,
a compact light source that is seen to rise above the inward extrapolated
surface brightness cusp at small radii.  In the present
sample we find 13/45 (29\%) of the core galaxies and 12/20 (60\%) of the
power-law galaxies to have nuclei.  Nuclei were recognized
by visual inspection of the residuals of the Nuker law fits to
the surface brightness profiles.  For the fainter nuclei in core
galaxies, the inner radial limits of the fits were often increased
to better isolate nuclei included in the initial fits.

Table \ref{tab:nuclei} lists the luminosities and colors of the
nuclei identified.  The luminosities were derived by integrating
the central residuals from the Nuker law fits.  This procedure presumes
that the Nuker law can be extrapolated inwards from the radial domain
over which the fit was performed.  The total fluxes of the nuclei may thus
have unknown systematic errors.  At the same time,
our nuclear luminosities for NGC 4278, 4365, and 4552 agree extremely
well with those of \citet{carollo}.  The color of the nucleus is
that of the central light source isolated from the hosting galaxy;
it is estimated from the Nuker law residuals in the $V$ band and the
$V-I$ profiles on the assumption that the $V-I$ color just outside the nuclei
gives the color of the background starlight.
The colors have been corrected for Galactic reddening.
The $\Delta(V-I)$ colors listed in Table \ref{tab:nuclei}
give the change in color of the nuclei relative to these fiducial values.
Note that a faint nucleus associated with a significant decrease
in the central total $V-I$ will thus be inferred to be extremely blue.

In general the nuclei are bluer than the background galaxy
starlight immediately outside the nuclei, although they are not
particularly blue in absolute terms.  Most nuclei are also spatially
unresolved (more compact than $\sim5$ pc); only the nuclei
in NGC 741 and 4365 appear to be resolved with scales of $\sim20$ pc
and $\sim10$ pc, respectively\footnote{\citet{carollo} noted the resolved
extent of the NGC 4365 nucleus, as well.}.
If the nuclei are star clusters, then they may comprise stars younger
or more metal-poor than those in the host galaxies.
We note that most of the nuclei are too luminous to be classic globular
clusters.  A better analog in color, luminosity, and physical
compactness would be the M33 nuclear star cluster \citep{l98}.
In passing, we note that NGC 5419 appears to have a double point source
at its center.  The brighter component is at the photocenter of the galaxy;
the second component appears to be weakly resolved.

The detection of nuclei in many power-law galaxies is not surprising
in light of previous work.  \citet{l95} identified nuclei in a large
fraction of their power-law galaxies (some of which are in the present sample);
\citet{rest} also found nuclei in a number of power-law galaxies.  The frequency
of nuclei in core galaxies is less well understood. \citet{l95} found
nuclei in only two of their core galaxies, NGC 6166, and the
brightest cluster galaxy (BCG) in A2052, both of which were known to be AGN.
However, the WFPC1 images were poorly suited to detecting faint nuclei.
\citet{carollo} were the first to use WFPC2 for this problem; they
only observed a few core galaxies, but found nuclei in most of them.
\citet{rest} also used WFPC2, but found nuclei in only one of their nine
core galaxies.  \citet{rav} found nuclei in 5/13 of their core
galaxies, consistent with the present detection rate.
Rest et al.\ adopted very conservative criteria for identifying nuclei, while
Ravindranath et al.\ always included a nuclear point source in
their Nuker law fits.  Lastly, \citet{laine} detected nuclei in $\approx15\%$
of their BCG sample, which mainly had core galaxies.  The BCG are in general
at much larger distances than the galaxies in the present sample and
observed in the $I$ band, both making the detection of nuclei difficult.
The present sample is also rich in core galaxies,
and the 29\%\ detection rate is probably a more
representative number than in previous samples.

There is too little information to determine whether nuclei are
star clusters or are AGNs; both phenomena are probably present.
\cite{rav} argued that at least 11/14 of their nuclei were associated with AGN.
In the present sample AGN are likely to explain NGC 1316 \citep{fab},
4552 \citep{cap}, and 7457 \citep{g03}.  Table \ref{tab:morph} lists what is
known about nuclear activity of the sample galaxies based
on the optical emission lines.
There is weak evidence that power-law galaxies with nuclei have
absorption-line spectra, while core galaxies with nuclei have
strong emission lines. However, exceptions in both directions exist;
in particular, a number of core galaxies with emission lines showed no nuclei
in the present surface photometry.  We thus cannot conclude
that low-level AGN, even if present, explain the all nuclear sources
seen in the broad-band {\it HST} images.

\subsection{Galaxies with Central Minima \label{sec:min}}

A basic structural paradigm that underlies the present structural analysis
of early type galaxies is that their central starlight distributions
can be described by concentric elliptical isophotes with surface
brightness increasing monotonically towards the center.
There is however, an increasingly large set of objects that depart
from this picture.  M31 \citep{l93} and NGC4486B \citep{l96} have double
nuclei, and \citet{l02} identify a number of galaxies that have local
{\it minima} in their brightness profiles at or near their centers.
In this section we discuss NGC 4073 and 4382, which also have central
surface brightness minima.  In the next section we discuss an additional
set of galaxies in which the location of maximum brightness within the
core is spatially offset from the photocenter of the core itself.

We speculate that a substantial majority if not all of these systems
may be cases where black holes were critical to either creating
or preserving the unusual structures over long times.
\citet{trem}, for example, explains the double nuclei in M31 (and by
implication, NGC 4486B) as the projected appearance of an eccentric
disk of starlight stabilized by a central black hole.
In the case of the galaxies with central minima,
\citet{l02} advanced two competing explanations
for the creation of such systems, both involving black holes.
The first is that a diffuse torus of starlight
has been added to a pre-existing core to create the appearance
of a central minimum interior to the torus.
This would occur during the final stages of merging in which the tidal field
of the black hole in the more massive galaxy disrupts the dense
center of a galaxy being cannibalized \citep{hbr2}.  A second, more exotic
idea is that stars have actually been ejected from the galaxy center
by the decay of a central black hole binary created in a merger.

\citet{l02} further noted that at least some of galaxies with central
minima may be related to the double nucleus systems.  Strictly speaking,
M31 and NGC 4486B have locations near their centers at which
a local minimum in surface brightness occurs, and could thus be included
in the class of galaxies with central minima.  Conversely, the
local minima in some of the \citet{l02} galaxies are bracketed by
structures that create the appearance of highly diffuse double nuclei.
Separately, the central minimum in NGC 3706 is clearly due to
a well defined central ring of stars that might create a double nucleus
if it were slightly more eccentric or were viewed from a different angle;
the central structure in NGC 3706 resembles
that of M31 in other ways.  In the present sample, the center of NGC 4073
resembles that of \citet{l02} galaxies NGC 4406 and 6876, while NGC 4382 is
yet another galaxy with a central minimum that resembles M31.
Lastly, we note that the galaxies with offset centers, which
we discuss in $\S\ref{sec:off},$ may be
poorly resolved examples of double nuclei or galaxies with central minima.

\subsubsection{NGC 4073}

The core galaxy NGC 4073 is the brightest galaxy in the poor cluster MKW4;
it is likely to have cannibalized other cluster members.
\citet{fisher} obtained a rotation curve and stellar
velocity dispersion profile.  The central few arcseconds of
the kinematic profiles reveal a $\sim10\%$ depression in velocity
dispersion and counter-rotation.  \citet{fisher} thus argue that NGC 4073
has a kinematically-decoupled core.

The envelope of NGC 4073 has a fairly high ellipticity of 0.4;
however, for $r<5'',$ the isophotes rapidly become nearly circular
and show a strong twist of over $90^\circ,$ as is evident in the profiles
shown in Figure \ref{fig:surf}, and in a contour plot of the galaxy
center (Figure \ref{fig:n4073_con}).  The maximum surface brightness
is reached in a ``ring'' at $r\approx0\asec16,$ with the
brightness falling by 0.14 mag at the center of the galaxy.
At larger radii the surface brightness profile (Figure \ref{fig:surf})
first {\it increases} in slope with decreasing radius for $r<1'',$
before abruptly flattening off just outside the ring.
Superficially, a high contrast
image of the galaxy center (Figure \ref{fig:n4073_im}, left) suggests that the
center is obscured by a dust patch; however, an archival NICMOS-2 image
(Figure \ref{fig:n4073_im}, right) (GTO 7820) in the $H$ band also shows a
central minimum, despite its lower
angular resolution.  We conclude that we are seeing the intrinsic
distribution of starlight rather than obscuration.

A close examination of the image and contour plot shows that the central
ring is not exactly symmetric, but is slightly brighter to one side.
A cut through the center at position angle $166^\circ$ shows this
clearly (Figure \ref{fig:n4073_cut}) by going through the brightest
part of the ring.  The maximum brightness along the cut occurs
at $0\asec18$ from the central minimum, slightly further
away from the center than the diametrically-opposed maximum, which
occurs at $0\asec15$ from the central minimum.  The implied eccentricity
of the ring is $e\sim0.07.$ The slightly asymmetric brightness distribution
could be explained by an eccentric torus of stars \citep{trem}.
However, the ring-radius corresponds to $\sim70$ pc, a considerably larger scale
than that of the central minima discussed in \citet{l02},
and one that is outside the nominal $\approx30$ pc sphere-of-influence of
a central black hole based on the velocity dispersion of NGC 4073
(Table \ref{tab:glob}) and the \citet{trem02} $M_\bullet-\sigma$
relationship.  If so, the stability of such a torus might therefore
be in question.

\subsubsection{NGC 4382}

The central structure of NGC 4382, a bright S0 core galaxy in the
Virgo cluster, somewhat resembles the double
nucleus of M31; here the central minimum is a ``valley'' between
two brightness peaks.  This might
result from the projected appearance of a much more eccentric torus
or stellar disk than that seen in NGC 4073.
As can be seen in Figures \ref{fig:n4382_im} and \ref{fig:n4382_con}
the isophotes become asymmetric for $r<0\asec5,$ rounding out
and becoming somewhat diffuse at position angle $\sim40^\circ.$
This side of the nucleus suggests a reduced-amplitude
version of the off-center P1 component of M31\footnote{P1 is the brighter,
more diffuse component of the M31 nucleus in the optical.
P2 is the less luminous peak, but it marks the center
of the galaxy and the location of the M31 nuclear black hole.
See \citet{l93} or \citet{l98} for details.};
indeed, a local surface brightness peak occurs at $\rm{PA}=64^\circ,$
separated by $0\asec14$ from the brightest portion of the nucleus,
which itself is within $0\asec01$ of the galaxy envelope photocenter.
Based on the velocity dispersion of NGC 4382 (Table \ref{tab:glob})
and the \citet{trem02} $M_\bullet-\sigma$ relationship, the
estimated sphere-of-influence of a central black hole is $\approx0\asec16.$
Again like M31, the isophotes become more eccentric and twist $>20^\circ$
within $r\sim1''$ as the nucleus is approached (Figure \ref{fig:surf}).
The position angle of the line connecting the two brightness peaks
is itself twisted away from isophotes of semimajor axes of only $\sim0\asec15.$
The isophotes on the side of the photocenter opposite the secondary peak
become highly eccentric, appearing as an almost linear extension
of the central nucleus; this in turn resembles the P2 component of M31.
Similar structure is seen in both the $V$ and $I$ bands; a color-ratio
image (Figure \ref{fig:n4382_im}) is flat (except for a slight radial gradient),
showing no evidence for dust or change in stellar population associated with
the unusual nuclear structure.

From other data, NGC 4382 appears to be an excellent candidate for a fairly
recent merger; the central structure noted here may be related to the unusual
spectral properties of the central regions detected from ground-based
observations.  NGC 4382 has the second-highest fine-structure measure in the
\citet{ss92} sample, who argue that the fine-structure
measure is indicative of the time and strength of the last merger.
\citet{ffi} obtained long-slit line-strength measures over the
central region of NGC 4382, finding it to have an exceptionally large
${\rm H}\beta$ absorption line-width of 2.6\AA; they argued that the central
population of NGC 4382 is as young as 3 Gyr.  They also confirm
that the galaxy becomes steadily {\it bluer} with decreasing
radius for $r>4'',$ as was first noted by \citet{bm87}.  The present $V-I$
color profile of NGC 4382 also shows a centrally blue gradient for
$r<10'';$ as discussed in
$\S\ref{sec:colgrad},$ nearly all galaxies have $V-I$ color gradients
that become redder at small radii.
We also note that NGC 4382 has an average $V-I$ color that is bluer than all
but one galaxy in the present sample (Table \ref{tab:vmi}).
We do, however, disagree with some of the finer
details of the NGC 4382 color distribution presented by \citet{ffi} ---
we find that the nucleus is essentially the same color as the
surrounding galaxy for $r<10'',$ and we do not find any evidence for
a central ring of particularly blue emission.
\citet{fish97} notes that the stellar velocity dispersion
profile of NGC 4382 decreases for $r<5'',$ which is confirmed in
SAURON observations \citep{sauron}.  High spatial-resolution
spectroscopy with the OASIS instrument shows that the
core within $r<2''$ counter-rotates with respect to the outer
envelope of the galaxy \citep{oasis}.

\subsection{Galaxies with Offset Centers \label{sec:off}}

In our data, the location of maximum surface brightness is typically concentric
with the surrounding galaxy, as measured by the center of the
isophotes on $\sim1''$ scales, with a median precision
of $0\asec005;$ 75\% of the sample has the peak brightness
concentric to within $0\asec01.$  In physical units, this means
that the galaxy centers are aligned at the sub-parsec level with the
isophotes on the scale of the break radii farther out.
Any lopsidedness or strong $m=1$ asymmetries on the scale of the
galaxy cores is therefore extremely rare (or of extremely low amplitude).
However, we identify five galaxies
(Table \ref{tab:morph}) in which the center of the surface brightness cusp
appears to be significantly displaced from the surrounding galaxy.

In NGC 507, the brightness peak is displaced $0\asec06$ from the
surrounding core; this is readily apparent in a contour plot
(Figure \ref{fig:n0507_con}) and in the residuals of the surface photometry
model (Figure \ref{fig:images}).
Examination of the image of NGC 1374 shows that its brightness
peak is clearly offset from its core (Figure \ref{fig:n1374_con}),
but the measured offset is only $0\asec02.$  The image of NGC 7619
looks normal, but the surface photometry model shows
a clear pattern associated with an offset peak, which is found to
be displaced from the core by $0\asec04.$
NGC 4291 and 5576 show similar
central offsets when the model is compared to the images.
In contrast, \citet{l95} identified
NGC 1700 as a possible candidate for an offset center, but the
improved imagery presented here shows no such effect.
The five galaxies identified here with offset centers are all core
galaxies, but the implied $\sim12\%$ fraction of core galaxies
with offset centers is too small to rule out similar behavior
in the power-law galaxies, which are more poorly represented
in the present sample.

It is now recognized that many early-type galaxies are triaxial,
rather than axisymmetric. The term ``triaxial" is usually taken to mean that
the galaxy is symmetric with respect to three orthogonal principal planes (the
$D_{2h}$ point symmetry group); if this symmetry is present then all
isophotes should be concentric. However, there is no known dynamical argument
that prohibits self-consistent stellar systems without this symmetry
(e.g., lopsided systems). The observation that the centers of most early-type
galaxies have concentric isophotes therefore shows that some aspect of the
galaxy-formation process strongly favors systems with $D_{2h}$ symmetry. In
the rare exceptions such as M31 \citep{l98}, the lopsidedness appears
to be localized within the sphere of influence of the central black hole,
which is natural since orbits do not precess in a Kepler potential, so a
lopsided orbit distribution is not axisymmetrized by phase mixing
\citep[e.g.,][]{peiris}.

The coincidence of the point of maximum surface brightness with the centroid
of the isophotes on larger scales also suggests that any massive black hole is
located at the bottom of the potential well due to the stellar mass
distribution to within a fraction of a parsec. Massive black holes are
normally expected to be close to the potential minimum because they rapidly
lose orbital energy due to dynamical friction; however, the black hole will
execute Brownian motion relative to the potential minimum due to gravitational
encounters with nearby stars.  The expected rms amplitude $r_{br}$ and velocity
$v_{br}$ of this motion have been discussed by many authors. \citet{bahcall}
argued that the black hole should be in energy equipartition with the
unbound stars surrounding it; for a hole of mass $M$ embedded in a
homogeneous core with core radius $r_c$ we then expect that
$r_{br}\sim r_c(m_*/M)^{1/2}$ and $v_{br}\sim \sigma(m_*/M)^{1/2}$
where $m_*$ is the typical stellar mass and $\sigma$ is the velocity
dispersion in the core. \citet{chatter} obtained a similar result,
which they confirmed using N-body simulations.  However, these results are
for galaxies with homogeneous cores, not the singular density
distributions that are found in both core and power-law galaxies.
There has been very little theoretical work on the expected value of
$r_{br}$ in galaxies with cusps.\footnote{\citet{lm04}
provide formulae for $v_{br}$ in the case where the galaxy density
profile is a power law, but do not discuss $r_b$.} However, it is likely that
the formula for a homogeneous core provides an approximate upper limit to
$r_b$ in a singular core, when the core radius of the homogeneous core is
replaced by the break radius of the singular core. For characteristic values
$r_c\sim 100$pc and $M\sim 10^8M_\odot$ we find $r_{br}\sim 0.01$pc,
too small to be resolved by {\it HST.}  Black holes in a binary
may also be displaced from the center.  The expected binary semi-major
axes are in the range 0.01 -- 10 pc \citep{yu}; binaries at the larger
end of this range may be within the resolution limit of {\it HST.}

The conclusion that Brownian wandering of a central black hole is not
responsible for the offset centers leaves us with the hypothesis
that these galaxies may be unresolved analogs to M31; this is suggested
by the highly local nature of the offset centers plus the increasingly
rich sample of early-type galaxies that have complex central structure.
\citet{nieto} noted that the M31 nucleus was displaced from the M31 bulge
based on ground-based observations, which had $\sim1$ pc limiting
resolution, prior to the {\it HST} discovery of the complex nuclear morphology
\citep{l93}.  If M31 were at the distance of Virgo, its double
nucleus would remain unresolved by {\it HST,} and would look
like an off-center galaxy.  The estimated sizes
of the black hole sphere-of-influence range from $0\asec11$ for NGC 5576
to $0\asec28$ for NGC 4291, significantly larger than the observed
central offsets for all five galaxies.  The structure of these
galaxies is thus not inconsistent with the \citet{trem} model for M31.

\section{Dust in the Centers of Early-Type Galaxies}

\subsection{The Phenomenology of Central Dust}

\subsubsection{Overall Dust Frequency}

It is already well-established that dust features are common
in {\it HST} images of the centers of early type galaxies
\citep{f94, jaffe, vdb, vdf, l95, carollo, ferrari, tomita, t01}.
In our own sample, dust is visible in 38 out of 77 galaxies.
This frequency of 49\% agrees well with that of previous studies
(\citealt{sg85}: 40\%; \citealt{vv88}: 23\%; \citealt{goud94a}:
41\%; \citealt{vdf}: 48\%; \citealt{ferrari}: 75\%;
\citealt{tomita}: 56\%; \citealt{t01}: 43\% (their unbiased sample)).
If we combine the samples of \citet{t01} and \citet{l95} with the
present sample (this combination largely encompasses all {\it HST} samples),
we find the overall frequency of dust to be 47\% in a total of 177 galaxies
(see Table \ref{tab:dustfreq} for details).  The sample of
\citet{l95} avoided known dust, whereas one quarter of the \citet{t01}
sample was strongly biased toward it; the present sample has no obvious
biases for dust.  The combined sample is large enough that the observed
dust fraction should approach a representative value for bright ellipticals.
Dust is about equally prevalent in both power-law galaxies and cores;
this agrees with frequencies seen in \citet{t01} and \citet{l95}.

\subsubsection{Dust Morphology and Radial Distribution}

Dust morphologies were determined by eye using the $4''$ postage stamps
shown in Figure \ref{fig:images}.  We should clarify that by ``dust''
in this paper we always mean clumpy dust that is visible as dark
patches in {\it HST} images.  A diffuse (and possibly more massive)
warm dust component may also be present, detected by IRAS
(e.g. \citealt{knapp}, \citealt{goud95}, \citealt{t01}),
but that is not the focus here.

Following \citet{t01}, we distinguish 
nuclear rings and disks (which we lump together as ``rings'')
from all other structures (termed ``non-rings'').  
Ring structures are presumably highly flattened, circular, 
and axi-symmetric \citep{vkdz}, implying ages that are at least a
few orbital times (an orbital time at 100 pc is a few million years).
The non-ring class is further sub-divided into ``spirals''
(with sheared dust patterns), and ``chaotics'' (random blobs).
A handful of transitional cases lie between the rings and non-rings.
This schema is motivated by the notion that we are seeing various stages of
a ``settling sequence" (e.g., \citealt{t01}; \citealt{vkdz}).
This idea is discussed further below, but it rests heavily on the fact
that non-ring dust {\it must} be in a dynamically short-lived
configuration.  The settling hypothesis says that it is falling
into the center to form nuclear rings.

The most striking result of the dust morphology census is that the ratio of
rings to non-rings is only 0.6, consistent across all three sub-samples.
This small ratio surprised us; if rings are the descendants of short-lived
non-rings, we expected to find a much larger ratio of rings to non-rings.
In fact, the opposite is true.  We return to this point below.

We studied larger model-subtracted images covering most of the PC1 CCD
to learn about the radial distribution of dust beyond $2'',$
which may be relevant to the settling model.\footnote{These larger-scale
images may be viewed at http://www.noao.edu/noao/staff/lauer/nuker.html.}
Dust concentration was graded qualitatively, with `0' meaning that
the dust is highly {\it un}concentrated --- dust is more prevalent
beyond the $4''$ postage stamp than in it, `1' means that dust is
visible both inside and outside of the postage stamp, and `2' means
that dust is seen only in the postage stamp itself.  The three dust
properties, strength (as already quantified in $\S\ref{sec:surf}$), 
morphology, and radial concentration, are presented in Table \ref{tab:morph}.

Only one of our 38 dusty galaxies (NGC 6876) has unconcentrated dust
(concentration class 0).  Thus, when dust appears in the envelope of the
galaxy, it essentially always in seen in the center, too.
The converse is not true.  Indeed, it is striking that galaxies with nuclear
rings are completely devoid of dust outside the ring: not one
dust cloud is to be found in these objects beyond the ring itself!

Overall, the radial pattern of dust suggests an {\it episodic} settling
model in which dust appears at times throughout the galaxy
and then falls to the center.  Galaxies with lots of dust would be
near the beginning of the cycle, when dust is still widely distributed.
Nuclear rings would be the endpoint of an episode, by which
time the rest of the galaxy would be dust free.  The fact that galaxies
with weaker dust also tend to be highly concentrated (Table \ref{tab:morph})
suggests that dust is being destroyed as it sinks to the center;
it falls in, leaving just a few weak shreds by the time it nears the nucleus.
A possible mechanism for dust destruction is sputtering by hot X-ray gas.
Survival of some but not all dust would then require that the timescale
for dust destruction by sputtering be comparable to the timescale for infall.
This is consistent with the dust-radiative cooling model of \citet{mb03}
(see also \citealt{sparks}).

\subsubsection{Dust and Nuclear Optical Emission} 

Many papers have found positive correlations between dust and nuclear optical
emission, extended optical emission, non-thermal radio activity, and far-IR
dust emission (e.g., \citealt{kim}; \citealt{shields}; \citealt{goud94a};
\citealt{vdf}; \citealt{vk99}; \citealt{martel99}; \citealt{tomita};
\citealt{t01}).  We decided to explore the relationship between dust and
{\it central} optical emission since that relation seemed particularly close
\citep{t01}.  Optical emission-line strengths were taken from the literature
(sources given in Table \ref{tab:morph}).  The emission is ranked qualitatively
into four grades in Table \ref{tab:morph}: 0 (no emission), 1 (weak emission),
2 (strong emission), and 3 (very strong emission).  Some papers measure
extended emission, but most are based on spectroscopic data and thus measure
only nuclear emission, which in our galaxies has line ratios characteristic
of active nuclei (Seyferts and LINERs).  Since extended and nuclear emission
correlate well \citep{kim, shields, goud94a, macchetto96, caon}, we assume
that any optical emission, however weak, signifies {\it nuclear activity.}

The presence of dust in the inner $4''$ of our galaxies correlates remarkably
well with the presence of emission: 96\% of galaxies without emission lack
dust, whereas 90\% of galaxies with emission have at least some dust.
Furthermore, the {\it amounts} of dust and emission also go together:
mean dust strength in weak emission galaxies is only 1.09, versus 1.85 in
strong-emission and very-strong-emission galaxies (see Table
\ref{tab:dustvsemiss}).  \citet{t01} also found a close connection
between dust and emission, though not quite so tight as in our sample.  
Notably, emission strength does not appear to depend on dust morphology;
as also found by \citet{t01}; rings and non-rings with the same dust strength
have qualitatively similar emission strengths.

This connection between dust and emission is important if it is assumed,
as here, that optical emission always signifies gas falling onto a black hole.
It then follows that visible dust actually {\it shows} us gas that is
about to fall onto the hole.  The distribution and motions of dust
become essential clues to the feeding of local AGNs, a point
that was also stressed by \citet{t01}.

\subsubsection{Dust and Galaxy Luminosity}

Figure \ref{fig:dust_mag} plots the histograms of dusty and
clean galaxies in Virgo as a function of absolute magnitude.
Dust is found in roughly half the galaxies brighter than $M_V = -21$ (which
is the normal frequency), whereas {\it all} 16 dimmer galaxies in Virgo are
dust free.  The probability of this happening by chance in 16 galaxies is only
$\sim 1.5 \times 10^{-5};$  if the dividing line were placed at $M_V = -20,$
it would still be only $2.4 \times 10^{-4}$.  Either
dust is intrinsically rare in faint ellipticals, or the cluster
environment discourages the accumulation of dust in small galaxies.  
Deciding between these hypotheses is problematic since the
only good selection of faint galaxies is in the Virgo cluster.

\subsection{Predictions of the Dust Settling Model}

The episodic settling model implies that we should to be able to
identify galaxies that are in different stages of settling.  A possible
sequence is offered in Figure \ref{fig:sequence}.  The transitional objects
that appear to be in the process of settling from a non-ring configuration
into a ring are particularly important.  Four (perhaps five) of these are in
our data, two of which are shown in Figure \ref{fig:sequence}.  \citet{vkdz}
note similar transitional structures in radio galaxies, which they call
``lanes." Lanes are geometrically distinct from rings and are more skewed with
respect to stellar principal planes, indicating transient configurations
that they suggest are in the process of settling.

If all these galaxies are really parts of a sequence --- non-rings, rings,
to empty --- then setting the lifetime of any one phase sets the lifetime
of the other two phases.  A good place to start is the non-ring phase.
Since non-ring dust cannot be in dynamical equilibrium,
a reasonable estimate of the lifetime is the local dynamical time.  The
orbital time at 1 kpc in a bright elliptical galaxy is
$\sim2 \times 10^7$ yr, and at the edge of a 100 pc nuclear ring
is ten times shorter.  Dust sputtering times (even with self-shielding)
are also of order $10^{7-8}$ years \citep{mb03}, which is consistent with
destruction of some fraction of the dust as it falls in.
Finally, once gas falls close to a rotating disk, the timescale
to settle into a principal plane of the gravitational potential is again
a few orbital times \citep{tsc82, sd88}.

The ratio of rings to non-rings can now be used to estimate the lifetime of
the ring phase.  Since that ratio is less than unity (0.6), the nuclear rings 
must be even shorter-lived!  \citet{jaffe96} model the kinematics of the nuclear
ring in NGC 4261, calling attention to the high turbulent velocity
near the center.  This central line-broadening is a common feature
of nuclear rings (e.g., NGC 7052, \citealt{vdb95}) and may be ubiquitous.
It implies high turbulent viscosity, say \citet{jaffe96}, which causes
the inner gas to flow inward and feed the black hole.  Their estimated
timescale for emptying out the inner ring is $\sim10^8$ yr, not far from
the timescale needed by the settling model.  Finally, rings should empty
from the inside out.
One might therefore expect a population of ``ghost" rings
that are the remnants of previously filled disks; four such
candidates are in our sample, two of which
are exceedingly faint (NGC 741, 2778, 3379, and 3608).

A further interesting observation is the finding by \citet{goud94b}
that dust grains are smaller in rings than in non-rings.  This
is consistent with the hypothesis that the grains in rings were subjected
to more sputtering by the hot gas, and are therefore older.

The final step is to deduce the length of time that galaxies remain
dust free, which sets the length of the whole cycle.  Since half of all
bright ellipticals are dust free, the length of the dust-free phase
must be of order that of the dusty phase, that is, a few times $\sim 10^7$
yr, and the length of the whole cycle is only $\sim 10^8$ yr.
The assumption of a settling sequence thus implies that {\it clumpy dust is
appearing and disappearing continually in ellipticals.}  This would
help to understand why there seems to be little correlation between
the presence of dust and merger remnants \citep{t01};
merger phenomena are comparatively long-lived and
would show little correlation with more rapid dust cycling.  The same
reason would explain why there is little correlation between mean
light-weighted stellar population ages and the presence
of dust (D. Forbes, private communication).

\subsection{Remaining Problems for the Dust Settling Model}

The ``dust settling'' hypothesis leaves at least three major questions
unanswered.  The first is the origin of the dust, whether
internal (shed by stellar winds) or external (accreted from
outside).   We find it striking that there is not the slightest
evidence, even in very dusty galaxies, for any extra starlight,
either distributed in clumps throughout the galaxy or in the neighborhood of
the dust clouds themselves.\footnote{NGC 1275 and 5128 provide
the only examples that we could find where this {\it does} happen;
neither galaxy is in our sample.  Both are conventionally explained
as mergers with an external gas-rich galaxy.}
Our subtracted models should be very
sensitive to clumpy starlight, yet no trace of stars is seen. Any dust
accreted from the outside must therefore be essentially {\it star free},
which excludes small dwarf galaxies as sources, or even tidal tails.  
We cannot think of any plausible external sources of dust that would
be completely star free.

In contrast, a known source of dust {\it within} the galaxies is M-star winds,
which are thought to have been detected in both the mid-IR and far-IR
\citep[e.g.][]{knapp, goud95, xil}.
However, this dust is injected uniformly throughout the galaxy and
is not expected to clump.  Perhaps there is a
thermal instability that causes the hot gas to cool into clumps,
bringing dust with it.  A thermal instability driven by dust cooling 
\citep{mb03} might play such a role.  Finding such a mechanism would
bolster the internal origin theory.

A second problem stems from the fact that the overall length for the settling
cycle seems uncomfortably short, only at most a few times $10^8$ yr.  If true,
this points strongly to an internal origin for the dust, as the
dynamical infall time from the edge of a galaxy is too long, more
like 1 Gyr.  Yet numerous studies have revealed galaxies in which
the distribution and motions of ionized gas and dust
seem {\it unrelated} to the motions of stars, and even counter-rotate
\citep[e.g.,][]{bert84, bert88, moll, kim, forbes, goud95, bert95,
bert98, corsini, caon, kaan}.
These data have been widely interpreted as incontrovertible
evidence for the {\it external} origin of dust, at least in certain key 
objects.  We have inspected our sample to identify
our best candidates for external dust origin, using the simple criteria
that the dust be extended and that its distribution over the galaxy
differ strongly from that of the underlying starlight.
Six objects meet these criteria (NGC 1316, 2768, 4026, 4589, 5018, and 7727).
Looking at them, it is indeed hard to imagine how the dust could have
come from any quasi-uniform process
{\it inside} these systems (see the large-scale model-subtracted
images noted in the footnote above). Five of these objects have strong dust,
and four also have strong optical emission, which would imply that
strong nuclear activity is triggered by external accretion in at least these
galaxies.  This would be consistent with the finding that two of the strongest
known LINERs (NGC 4278 and NGC 1052) have external 
H I clouds in their vicinity \citep{raim, vang}
as does NGC 5018 \citep{kim}, another AGN in our sample.
On the other hand, \citet{t01} searched for relationships between
dust strength and external environments, finding no connection between dust
and cluster versus group membership, or the distance to the
nearest neighboring galaxy.

New data on the interstellar medium in ellipticals may help to solve this
problem by {\it overturning} the old prediction that gas and star kinematics
must agree.  A consensus seems to be building that the hot
interstellar medium (ISM) in ellipticals is highly turbulent and is being
stirred periodically by nuclear activity \citep{enss, bill}.
Stirring of the ISM would strongly perturb warm ionized gas and dust,
and would explain the many examples of highly irregular, disorganized
velocity fields that are detected in warm gas (e.g., \citealt{caon}).
Any dust and gas thrown out to the periphery of galaxies by stirring
could be further perturbed by the passage of neighboring
galaxies or the pressure of hot intracluster winds.  In such a way,
the motions of dust, gas, and stars might become strongly decoupled,
giving the appearance of an external origin for clumpy gas and dust
when none is required.

The most important problem with the cycling picture is a clear idea
of how all the stages fit together and what drives the whole cycle.
Why does dust begin to condense into clouds in the first place?
Why does the black hole accrete only dust-related gas and not other gas?
What stops dust from condensing?  If it is feedback from the black hole,
what exactly is its nature and how does it interface with dust and
dust-related gas?  These questions are basic to the cycling model
but currently lack answers.

We conclude by highlighting the beautiful but puzzling
object NGC 3607 (Figure \ref{fig:n3607}).  This galaxy contains a large,
dusty outer disk that must be dynamically old because of its extraordinary
symmetry and tightly wrapped arms.  However, this outer disk appears
to transition rapidly but smoothly at the center to a second gas disk
that is {\it perpendicular} to the first and is seen nearly edge on!
This inclined disk seems to be settling onto an inclined nuclear ring.
How can this happen?  A facile explanation would invoke external
infall of gas directly into the galaxy center, yet no disturbance in 
the outer dust or other sign of an interaction is seen.  Thanks to its 
many regularities but schizophrenic nature, NGC 3607 may be a
Rosetta Stone that will reward further study.

\section{Summary}

We have examined the central structure of early-type galaxies;
a component of this analysis has included searching for
morphological differences between core and power-law galaxies.
We conclude by summarizing what this analysis might say
about the formation of central structure in the context of hierarchical
merging of galaxies with central black holes; in \citet{f97}
we argued that cores and power-laws are different outcomes of this process.
This discussion is largely speculative  --- while modern merger simulations are
able to study the gross effects of mergers on central structure,
there are few firm theoretical predictions of the phenomena
considered in the analysis sections.

The picture presented by \citet{f97} is that power-law
galaxies were formed by a process in which central {\it gaseous} dissipation
and attendant star formation took place, while core galaxies
were generated in the mergers of galaxies largely {\it free of gas.}
The first generation of core galaxies would be made by merging power-laws,
which themselves would harbor central black holes formed
in the initial collapse of the primordial galaxy.
The cores were formed in the final stage of the merger
when the central black holes of the merging galaxies formed a
binary that ejected stars from the center as it hardened.
We can distill this scenario to be simply
that power-law centers are sites of dissipation and concentration,
while cores are sites of mixing and scattering.

In the case of color gradients, this picture leads us to expect
power-law galaxies to have stronger gradients than core galaxies.  As
star formation progresses following a merger or initial collapse
of gas-rich systems, the inventory of gas is simultaneously slowly
depleted and centrally concentrated, while becoming increasingly
metal-enriched.  The stellar density profile would reflect both the
concentration of gas over time and the metal-enrichment history.
Merging a then gas-free power-law galaxy with a similar system
would not completely erase any pre-existing color gradients,
but some dilution of the gradient might be expected.  Strong
mixing of stars interior to the break radius associated with the
formation of a core, however, might cause significant flattening
of the gradient at small radii.  The present data do show that power-law
galaxies on average have steeper color gradients than do core galaxies,
although the difference is small and the scatter about the mean gradient for
both types is large.  Whether or not the color gradients in core
galaxies flatten further interior to the break radius is unclear.
There are some core galaxies that do appear to show this effect,
but the majority do not.  Unfortunately, in absolute terms the envelope
gradients in core galaxies are small to begin with, and
it is thus especially difficult to detect even smaller gradients
over a limited radial range.  We conclude that color gradients are broadly
consistent with the formation of cores by binary black holes with the
caveat that the effects seen are modest in absolute terms.

The discovery that core galaxies twist interior to their break radii
at first glance also appears to support the formation of cores
by binary black holes.  If the shape of the core is related to
the angular momentum vector of the binary black hole, which itself
may have little to do with the principal axes of the merged
galaxy, then it is easy to imagine that twists occur as one
transitions from the outer regions of the core into the inner
region dominated by the black hole.  However, our analysis shows that the
absolute physical significance of the twists must also be considered.
Since the inner twists of both cores and power-laws are about the
same in terms of the strength of the photometric perturbation
required to twist the galaxy, it is not possible to argue that
the {\it angularly} larger twists in core galaxies are a consequence
of their particular formation mechanism.

The interpretation of nuclei in early-type galaxies
depends on whether the nuclei are the non-stellar emission
of low-level AGN, or are star clusters of the sort seen in M33.
The existence of nuclear star clusters in cores raises a few questions.
In \citet{f97} we argued that the survival of cores, once created, against
infalling material in subsequent mergers requires the central black hole to
serve as a ``guardian'' that tidally disrupts any dense stellar aggregates
that reach the center, a notion that has been validated by N-body
simulations \citep{hbr2}.  It thus appears that nuclear star clusters
cannot be pre-existing systems that have arrived at the
center through dynamical friction.  The likely alternative is that
such nuclei were created {\it in situ} by the central accretion of
small amounts of gas with subsequent star formation.  This scenario
may be supported by the bluer colors of the nuclei if they
indicate that the nuclei are younger than the rest of the
central stellar population.  The remaining issue is to understand
how converting any infalling gas into stars competes with accretion
into the black hole, itself.  The same problem exists in the center
of our own Galaxy, where a very young star cluster surrounds a
central black hole (see the review by \citealt{ms}).
We conclude that while formation of nuclear clusters is an interesting problem,
their existence does not offer any obvious insight into the formation
of core versus power-law galaxies.  This conclusion also applies to AGN,
which in this sample offer little more than a ratification of the already
standard notion that black holes are likely to exist at the centers
of both types of galaxies.

The discovery of two more galaxies that have central minima in their
brightness distributions in addition to the galaxies discussed in \citet{l02}
implies that $\sim5\%$ of core galaxies have such structures.
As noted in \citet{l02},
there are two competing ideas for the formation of these objects.
In the first scenario, central minima might
be explained by diffuse tori of stars added to a pre-existing core.
The tori result from the
tidal disruption of dense stellar aggregates transported to the
center of the galaxy by dynamical friction following a merger;
again, the black hole prevents the core once established from being
filled in by latter mergers.
In a second and more speculative scenario, the central minima
are a direct consequence of stars ejected from the galaxy center by
a binary black hole.  It may be noteworthy
that NGC 4073 and a number of the galaxies discussed in \citet{l02}
have relatively high luminosities.  In the discussion in that paper,
we advanced the hypothesis that the mergers of two galaxies
each with pre-existing cores, as would occur at the high-mass
ends of a merging hierarchy, might be the necessary precondition
for generating a merged galaxy in which the core was actually
evacuated by the binary black hole.
Under this picture, galaxies with central minima would be vivid
examples of the action of binary black holes on central structure;
however, present theoretical and observational treatment of this problem
leaves this as an unproven hypothesis.

Likewise, we presently do not know what bearing galaxies with offset centers
have on understanding the formation of central structure.  If
these systems are indeed unresolved analogs of the M31 nucleus, then
their role may be more to show how nuclear black holes can stabilize
non-axisymmetric structures in galaxy centers.

In \citet{f97} we noted that power-law galaxies have disky outer isophotes;
the conclusion here that all inclined power-law galaxies
have visible stellar disks, with the implication that disks
exist in {\it all} power-law galaxies, supports the dissipative
formation scenario for power-law galaxies.  Conversely, the
corresponding lack of disks in all but the lowest-luminosity core
galaxies supports their formation in dissipationless
gas-free processes.  A transition zone in galaxy luminosity where
the dissipation and dissipationless stellar
processes compete in mergers may explain the disks in the lower-luminosity
galaxies with cores.  Lastly, the rounder shapes of the inner isophotes of
core galaxies as compared to power-law galaxies may be a direct consequence
of core formation by a binary black hole on the presumption that scattering
of stars by the binary is most likely to be nearly isotropic.

In the final analysis section, we have reasoned that the high
prevalence of clumpy dust in non-equilibrium conditions
means that it is generated in processes interior
to the galaxies, themselves.  We have further argued that dust clouds
once created settle into the center, and are eventual destroyed or
processed by the galaxy on time scales of $\sim10^8$ yr.  The
continued existence of dust clouds in a substantial fraction of
early-type galaxies suggests that dust clouds are formed and
destroyed in a cyclical process.  The strong link between the
presence of dust and nuclear emission argues that one outcome
of this process would be episodic delivery of matter to the centers
of the galaxies; understanding the secular evolution of cores may have
to take this into account.  At the same time we see no
properties of present-day dust that bear one way or another on the
basic formation mechanism of cores versus power-laws.

In conclusion, it appears that the diverse morphological measures
derived for galaxies here give at most weak evidence for differences
between core and power-law galaxies that might be taken as support
for the binary black hole core formation mechanism.
Dust is formed equally in both types, and the nuclei offer no apparent
insight into this problem.  Understanding the galaxies with
central minima may offer a useful path for understanding core formation,
but presently the mechanisms for forming central minima are too
uncertain for forming conclusions.  Position-angle twists, when
properly normalized, seem equally strong in both types.  Cores
do have shallower color gradients, as expected, but the effect is weak.
The strongest evidence comes from the rounder inner isophotes
of cores and, more important, the high frequency of disks in
the centers of power-law galaxies and (to a lesser degree) in the
fainter core galaxies.  Overall the evidence is consistent
with the binary black hole core formation mechanism, but it
is not a strong endorsement.

The real value of our data, we feel, is not in current
tests of core formation, but in providing a
rich set of observations that can be used to confront future models.
Any successful model will have to match the large suite of
quantitative data produced in this paper.
Finally, we note that this paper does not consider the fundamental
radial distribution of starlight in galaxies, which distinguishes
cores from power-laws in the first place.  This is the topic
that taken up in Paper VI, where we demonstrate a strong dichotomy
between the two types, validating the assumption of two classes in this paper.

\acknowledgments

This research was supported in part by several grants provided through STScI
associated with GO programs 5512, 6099, 6587, 7388, 8591, 9106, and 9107.
Our team meetings were generously hosted
by the National Optical Astronomy Observatory, the Observatories
of the Carnegie Institution of Washington, the Aspen Center
for Physics, and the Leiden Observatory.  We thank Dr. Barbara Ryden
for useful conversations on isophote twists.

\clearpage

\begin{figure}
\caption{{\it Available as standalone JPEG figures.}
The deconvolved central $4''\times4''$ ($88\times88$ PC1 pixels)
regions of the galaxies are shown.  The stretch is arbitrary and is adjusted to
maximize image contrast over the portion of each galaxy shown.  The image
orientation is determined by how they were acquired in the PC1 CCD;
arrows indicate the direction to north.  Galaxies that were double-sampled
(see $\S\ref{sec:red}$) have been binned
to the original pixel scale for this figure.  For galaxies observed in
multiple bandpasses, the F555W image has been shown.  The panel to
the right of each central image is the same region divided by a galaxy model
reconstructed from the surface photometry; the contrast stretch is set
to $\pm50\%$ black to white.  Galaxies with obscured centers are
shown in Figure \ref{fig:obs_im}.}
\label{fig:images}
\end{figure}

\begin{figure}
\caption{{\it Available as a standalone JPEG figure.}
The central $4''\times4''$ ($88\times88$ PC1 pixels)
regions of the galaxies with dust-obscured centers are shown.  These are
objects for which no surface-brightness models could be derived.  The image
characteristics are otherwise the same as for Figure \ref{fig:images}.}
\label{fig:obs_im}
\end{figure}

\begin{figure}
\caption{{\it Available as standalone GIF figures.}
Surface photometry is shown for the unobscured galaxies.  The top
panel shows the surface brightness profile as a function of semi-major axis.
Photometry is shown for the F555W filter, except for the few galaxies that were
observed in this band.  No Galactic absorption corrections have been applied.
The solid line is a Nuker law (equation \ref{eqn:nuker}) fitted to the
brightness profile; the extent of the line shows the radial range of the fit.
The trace at the bottom of the panel shows the residuals from the Nuker law fit,
referenced to the scale on the right side of the panel in magnitudes.
The scale at the top of the panel is in parsecs, based on the distances given
in Table \ref{tab:glob}.  The next two thin panels for each galaxy give
isophote position angle and ellipticity ($1-{b\over a}$).
The range of the position angle plot varies depending on the range of
isophote position angles observed; in some cases the minimum PA is
negative to suppress abrupt $180^\circ$ jumps.  For the core galaxies,
a dashed line shows fits of the PA as a function of radius to model
described in $S\ref{sec:twist}.$ The bottom panel shows the $V-I$ profile
with fitted gradient for those galaxies that have color information.}
\label{fig:surf}
\end{figure}

\clearpage

\begin{deluxetable}{lcclcl}
\tablecolumns{6}
\tablewidth{0pt}
\tablecaption{Observational Summary}
\tablehead{\colhead{Galaxy}&\colhead{Program}
&\colhead{Filter}&\colhead{Exposure (s)}&\colhead{Sky}&\colhead{Notes}}
\startdata
NGC 0507     &6587&F555W&$2\times800$             &22.44&                    \\
NGC 0584     &6099&F814W&$6\times260$             &21.07&                    \\
            &     &F555W&$4\times350$             &21.97&                    \\
NGC 0596     &6587&F555W&$4\times400$             &22.36& Double-sampled\\
NGC 0741     &6587&F555W&$2\times400$             &21.84&                    \\
NGC 0821     &6099&F814W&$6\times260$             &21.28& \citet{g03}\\
            &     &F555W&$4\times350$             &22.18&                    \\
NGC 1016     &6587&F555W&$2\times800$             &22.37&                    \\
NGC 1023     &6099&F814W&$2\times180$             &21.48& \citet{bower}\\
            &     &F555W&$2\times180$             &22.38&                    \\
             &6587&F555W&$5\times260$             &21.40& Double-sampled\\
NGC 1316     &5990&F814W&$600+260$                &20.69&                    \\
NGC 1374     &5446&F606W&$2\times80$              &22.57&                    \\
NGC 1399     &8214&F606W&$8\times500$             &21.75& Double-sampled\\
NGC 1426     &6587&F555W&$4\times400$             &22.22& Double-sampled\\
NGC 1427     &5454&F814W&$2\times230$             &21.40& \citet{carollo}\\
            &     &F555W&$2\times500$             &22.50&                    \\
NGC 1439     &5454&F814W&$2\times230$             &21.67& \citet{carollo}\\
            &     &F555W&$2\times500$             &22.57&                    \\
NGC 1700     &5454&F814W&$2\times230$             &21.05& \citet{carollo}\\
            &     &F555W&$2\times500$             &21.95&                    \\
             &6587&F555W&$4\times400$             &22.23& Double-sampled\\
NGC 2300     &6099&F814W&$6\times230$             &21.56&                    \\
            &     &F555W&$4\times350$             &22.46&                    \\
NGC 2434     &5943&F450W&$1100+800$               &23.21&                    \\
            &     &F814W&$600+400$                &21.54&                    \\
            &     &F555W&$700+600$                &22.26&                    \\
NGC 2768     &6587&F555W&$4\times100+2\times500$  &21.99& Double-sampled; obscured\\
NGC 2778     &6099&F814W&$6\times260$             &21.73& \citet{g03}\\
            &     &F555W&$4\times350$             &22.60&                    \\
NGC 2974     &6587&F555W&$4\times100+2\times400$  &22.94& Double-sampled\\
NGC 3115     &5512&F814W&$3\times350$             &20.83& \citet{k96}        \\
            &     &F555W&$3\times350$             &21.65&                    \\
NGC 3377     &5512&F814W&$3\times350$             &21.38& \citet{g03}\\
            &     &F555W&$3\times350$             &22.29&                    \\
NGC 3379     &5512&F814W&$3\times400$             &21.40& \citet{g3379}\\
            &     &F555W&$3\times500$             &22.30&                    \\
NGC 3384     &5512&F814W&$3\times350$             &22.35& \citet{g03}\\
            &     &F555W&$3\times350$             &22.35&                    \\
NGC 3557     &6587&F555W&$2\times900$             &\dots& Obscured           \\
NGC 3585     &6587&F555W&$4\times100+2\times400$  &21.75& Double-sampled\\
NGC 3607     &5999&F814W&$160+160$                &21.24&                    \\
            &     &F555W&160                      &22.21&                    \\
NGC 3608     &5454&F814W&$2\times230$             &21.64& \citet{carollo}\\
            &     &F555W&$2\times500$             &22.54&                    \\
NGC 3610     &6587&F555W&$4\times100$             &22.59& Double-sampled\\
NGC 3640     &6587&F555W&$4\times100+2\times400$  &21.82& Double-sampled\\
NGC 3706     &6587&F555W&$4\times100+2\times500$  &21.82& Double-sampled; \citet{l02}\\
NGC 3842     &6587&F555W&$2\times800$             &22.54&                    \\
NGC 3945     &6633&F450W&$1100+900$               &22.36&                    \\
            &     &F814W&$500+350$                &21.77&                    \\
            &     &F555W&$700+500$                &22.68&                    \\
NGC 4026     &9107&F814W&$8\times200$             &21.57& Double-sampled\\
            &     &F555W&$8\times200$             &22.83& Double-sampled\\
NGC 4073     &6587&F555W&$2\times800$             &22.30&                    \\
NGC 4125     &6587&F555W&$4\times100+2\times500$  &\dots& Double-sampled; obscured\\
NGC 4278     &5454&F814W&$2\times230$             &21.86& \citet{carollo}\\
            &     &F555W&$2\times500$             &22.76&                    \\
NGC 4291     &6099&F814W&$6\times260$             &21.98& \citet{g03}\\
            &     &F555W&$4\times350$             &22.88&                    \\
NGC 4365     &5454&F814W&$2\times230$             &21.43& \citet{carollo}\\
            &     &F555W&$2\times500$             &22.33&                    \\
NGC 4382     &7468&F814W&$4\times800$             &21.07&                    \\
            &     &F555W&$4\times700$             &22.07&                    \\
NGC 4406     &5512&F814W&$3\times500$             &21.40& \citet{l02}\\
            &     &F555W&$3\times500$             &22.30&                    \\
NGC 4458     &6587&F814W&$2\times800$             &21.55&                    \\
            &     &F555W&$2\times400+199$         &22.49&                    \\
NGC 4472     &5236&F814W&$2\times900$             &21.40&                    \\
            &     &F555W&$2\times900$             &22.30&                    \\
NGC 4473     &6099&F814W&$4\times500$             &21.38& \citet{g03}\\
            &     &F555W&$3\times600$             &22.32&                    \\
NGC 4478     &6587&F555W&$4\times400$             &22.12& Double-sampled\\
NGC 4486B    &6099&F814W&$4\times500$             &\dots& \citet{l96}\\
            &     &F555W&$3\times600$             &\dots&                    \\
NGC 4494     &5454&F814W&$2\times230$             &21.70& \citet{carollo}\\
            &     &F555W&$2\times500$             &22.60&                    \\
NGC 4552     &6099&F814W&$3\times500$             &21.17& \citet{carollo}\\
            &     &F555W&$4\times600$             &22.15&                    \\
NGC 4589     &5454&F814W&$2\times230$             &22.06& \citet{carollo}\\
            &     &F555W&$2\times500$             &22.96&                    \\
NGC 4621     &5512&F814W&$3\times350$             &21.19&                    \\
            &     &F555W&$3\times350$             &22.09&                    \\
NGC 4649     &6286&F814W&$1300+1200$              &21.40& \citet{g03}\\
            &     &F555W&$1100+1000$              &22.30&                    \\
NGC 4660     &5512&F814W&$4\times200$             &20.89&                    \\
            &     &F555W&$4\times230$             &21.85&                    \\
NGC 4709     &6587&F555W&$2\times900$             &21.98&                    \\
NGC 4786     &6587&F555W&$4\times100$             &\dots& Double-sampled; obscured\\
NGC 4936     &6587&F555W&$2\times800$             &\dots& Obscured           \\
NGC 5018     &6587&F555W&$4\times100$             &21.94& Double-sampled; obscured\\
NGC 5061     &6587&F555W&$4\times100+2\times400$  &22.48& Double-sampled\\
NGC 5322     &5454&F814W&$2\times230$             &\dots& Obscured; \citet{carollo}\\
            &     &F555W&$2\times500$             &\dots&                    \\
NGC 5419     &6587&F555W&$2\times800$             &22.23&                    \\
NGC 5557     &6587&F555W&$4\times100+2\times500$  &22.78& Double-sampled\\
NGC 5576     &9107&F814W&$4\times400$             &21.60& Double-sampled\\
            &     &F555W&$4\times400$             &22.52& Double-sampled\\
NGC 5813     &5454&F814W&$2\times230$             &21.51& \citet{carollo}\\
            &     &F555W&$2\times500$             &22.41&                    \\
NGC 5845     &6099&F814W&$4\times260$             &\dots& Obscured           \\
            &     &F555W&$3\times400$             &\dots&                    \\
NGC 5982     &5454&F814W&$2\times230$             &22.15& \citet{carollo}\\
            &     &F555W&$2\times500$             &23.15&                    \\
NGC 6776     &6587&F555W&$2\times900$             &\dots& Obscured           \\
NGC 6849     &6587&F555W&$2\times900$             &22.56&                    \\
NGC 6876     &6587&F814W&$2\times1400+1300$       &\dots& \citet{l02}\\
            &     &F555W&$2\times1400+1300$       &\dots&                    \\
NGC 7213     &9107&F814W&$8\times50$              &21.83& Double-sampled\\
            &     &F555W&$8\times50$              &22.84& Double-sampled\\
NGC 7457     &5512&F814W&$3\times350$             &21.93& \citet{g03}\\
            &     &F555W&$3\times350$             &22.87&                    \\
NGC 7619     &6554&F814W&$1300+900$               &21.59&                    \\
            &     &F555W&$1300+900$               &22.49&                    \\
NGC 7626     &5454&F814W&$2\times230$             &\dots& Obscured; \citet{carollo}\\
            &     &F555W&$2\times500$             &\dots&                    \\
NGC 7727     &7468&F814W&$4\times800$             &20.61&                    \\
            &     &F555W&$3\times700+533$         &21.50&                    \\
NGC 7785     &6587&F555W&$4\times100+2\times400$  &22.43& Double-sampled\\
IC 1459      &5454&F814W&$2\times230$             &21.59& \citet{carollo}\\
            &     &F555W&$2\times500$             &22.49&                    \\
IC 3370      &6587&F555W&$4\times100+2\times500$  &\dots& Double-sampled; obscured\\
IC 4296      &6587&F555W&$2\times800$             &\dots& Obscured           \\
IC 4329      &6587&F555W&$2\times800$             &22.56&                    \\
ESO 0462-015 &6587&F555W&$2\times800$             &22.45&                    \\
\enddata
\label{tab:obs}
\tablecomments{``Double-sampled" refers to galaxies for which dithered
images were used to construct an image with $2\times2$ pixel subsampling.
``Obscured" means that the center of the galaxy was either not visible due to
dust obscuration, or was so strongly affected by dust that
accurate surface photometry could not be extracted. Sky values are in
magnitudes per square arcsec. The references listed refer to earlier
publication of surface photometry derived from the specific images listed.}
\end{deluxetable}

\begin{deluxetable}{llrccr}
\tablecolumns{6}
\tablewidth{0pt}
\tablecaption{Global Parameters}
\tablehead{\colhead{Galaxy}&\colhead{Morph}
&\colhead{D (Mpc)}&\colhead{Ref}&\colhead{$M_V$}&\colhead{$\sigma$ (km/s)}}
\startdata
NGC 0507       &S0    & 63.8&4&$-$23.38&329\\
NGC 0584       &E     & 22.1&1&$-$21.39&225\\
NGC 0596       &E     & 22.1&1&$-$20.99&164\\
NGC 0741       &E     & 70.9&4&$-$23.20&293\\
NGC 0821       &E     & 25.5&1&$-$21.39&209\\
NGC 1016       &E     & 75.9&3&$-$23.03&279\\
NGC 1023       &S0    & 12.1&1&$-$20.51&212\\
NGC 1316       &E     & 22.7&1&$-$23.34&250\\
NGC 1374       &E     & 20.9&1&$-$20.75&207\\
NGC 1399       &E     & 21.1&1&$-$22.07&359\\
NGC 1426       &E     & 26.5&1&$-$20.80&153\\
NGC 1427       &E     & 21.0&1&$-$20.78&170\\
NGC 1439       &E     & 26.5&1&$-$20.98&159\\
NGC 1700       &E     & 40.6&2&$-$21.94&234\\
NGC 2300       &S0    & 30.4&4&$-$21.72&252\\
NGC 2434       &E     & 22.8&1&$-$22.01&229\\
NGC 2768       &E     & 23.3&1&$-$21.90&194\\
NGC 2778       &E     & 24.2&1&$-$19.52&166\\
NGC 2974       &E     & 22.7&1&$-$21.33&143\\
NGC 3115       &S0-   & 10.2&1&$-$21.17&271\\
NGC 3377       &E     & 11.7&1&$-$20.04&143\\
NGC 3379       &E     & 11.7&1&$-$20.90&221\\
NGC 3384       &S0-   & 11.7&1&$-$19.88&158\\
NGC 3557       &E     & 48.3&1&$-$23.23&265\\
NGC 3585       &E     & 21.2&1&$-$22.01&216\\
NGC 3607       &E     & 10.9&1&$-$20.64&220\\
NGC 3608       &E     & 23.0&1&$-$21.11&195\\
NGC 3610       &E     & 22.6&1&$-$20.88&167\\
NGC 3640       &E     & 28.3&1&$-$21.97&184\\
NGC 3706       &S0-   & 46.9&4&$-$22.49&302\\
NGC 3842       &E (BGC) & 97.0&1&$-$23.22&316\\
NGC 3945       &S0    & 19.9&3&$-$21.14&176\\
NGC 4026       &S0    & 15.6&1&$-$20.32&193\\
NGC 4073       &E     & 92.2&4&$-$23.62&266\\
NGC 4125       &E     & 25.2&1&$-$22.34&233\\
NGC 4278       &E     & 16.7&1&$-$21.05&251\\
NGC 4291       &E     & 25.0&1&$-$20.56&284\\
NGC 4365       &E     & 21.6&1&$-$22.02&269\\
NGC 4382       &S0    & 17.9&1&$-$22.28&184\\
NGC 4406       &E     & 17.9&1&$-$22.28&246\\
NGC 4458       &E     & 17.9&1&$-$19.32&101\\
NGC 4472       &E     & 17.9&1&$-$22.92&303\\
NGC 4473       &E     & 17.9&1&$-$20.99&179\\
NGC 4478       &E     & 17.9&1&$-$19.98&144\\
NGC 4486B      &cE    & 17.9&1&$-$17.91&178\\
NGC 4494       &E     & 17.9&1&$-$21.48&155\\
NGC 4552       &E     & 17.9&1&$-$21.39&263\\
NGC 4589       &E     & 25.0&1&$-$21.33&225\\
NGC 4621       &E     & 17.9&1&$-$21.61&237\\
NGC 4649       &E     & 17.9&1&$-$22.49&343\\
NGC 4660       &E     & 17.9&1&$-$20.09&191\\
NGC 4709       &E     & 37.0&1&$-$22.31&239\\
NGC 4786       &E+p   & 70.7&4&$-$22.58&295\\
NGC 4936       &E     & 50.1&4&$-$23.26&278\\
NGC 5018       &E     & 42.6&3&$-$22.61&212\\
NGC 5061       &E     & 26.9&4&$-$21.94&194\\
NGC 5322       &E     & 33.0&1&$-$22.40&234\\
NGC 5419       &E     & 62.6&4&$-$23.43&315\\
NGC 5557       &E     & 52.7&4&$-$22.61&253\\
NGC 5576       &E     & 27.1&1&$-$21.67&188\\
NGC 5813       &E     & 28.7&1&$-$21.84&230\\
NGC 5845       &E     & 28.7&1&$-$19.84&241\\
NGC 5982       &E     & 40.5&4&$-$22.01&250\\
NGC 6776       &E+p   & 77.6&4&$-$22.80&209\\
NGC 6849       &S0-   & 84.0&4&$-$23.32&207\\
NGC 6876       &E     & 57.6&4&$-$23.74&230\\
NGC 7213       &Sa    & 22.7&3&$-$21.51&185\\
NGC 7457       &S0-   & 14.0&1&$-$18.63& 78\\
NGC 7619       &E     & 56.1&1&$-$22.93&319\\
NGC 7626       &E     & 56.0&1&$-$22.49&269\\
NGC 7727       &Sa    & 21.6&3&$-$21.33&200\\
NGC 7785       &E     & 49.9&4&$-$22.14&253\\
IC 1459        &E     & 30.9&1&$-$22.56&311\\
IC 3370        &E     & 28.3&1&$-$21.55&202\\
IC 4296        &E (BCG) &45.4&2&$-$23.32&340\\
IC 4329        &E (BCG) & 56.7&2&$-$23.70&270\\
ESO 462-15    &E     & 80.3&4&$-$22.88&289\\
\enddata
\label{tab:glob}
\tablecomments{Morphological classifications are from the NASA/IPAC
Extragalactic Database;
BGC means that the galaxy is a brightest cluster galaxy.
Distance reference codes: (1) \citet{ton} SBF distances
scaled to $H_0=70$ km s$^{-1}$ Mpc$^{-1}$; (2) distance from
\citet{f97} scaled to $H_0=70;$ (3) \citet{f89} group distance
scaled to $H_0=70;$ and (4) distance based on CMB velocity of galaxy or group.
Velocity dispersions are provided by the \citet{ps} compendium
of all published velocity dispersions.}
\end{deluxetable}

\begin{deluxetable}{lcccccclll}
\tablecolumns{10}
\tablewidth{0pt}
\tablecaption{Central Properties}
\tablehead{\colhead{ }&\multicolumn{3}{c}{\underbar{\hbox to 90pt{\hfill Dust
\hfill}}}&
\colhead{Stellar}&\colhead{ }&\colhead{ }&\colhead{Spec}&\colhead{Emiss}&
\colhead{ } \\
\colhead{Galaxy}&\colhead{Str}&\colhead{Morph}&\colhead{Conc}&
\colhead{Disk}&\colhead{Nuc}&\colhead{Pec}&\colhead{Type}&\colhead{Class}&
\colhead{Ref}}
\startdata
NGC 0507     & 0.0 &     &    &               &\phantom{$\times$} & Offset   & a & 0  & 1,3,8             \\
NGC 0584     & 0.5 & c   & 2  &               &\phantom{$\times$} &          & E:     & 1: & 3,5,6,8,12    \\
NGC 0596     & 0.0 &     &    &               &$\times$           &          & a & 0  & 4,5,8,9,12       \\
NGC 0741     & 0.0 & nr  & 2  &               &$\times$           &          & E & 1  & 5,13              \\
NGC 0821     & 0.0 &     &    & $\times$      &\phantom{$\times$} &          & a & 0  & 1,3,4,5,8     \\
NGC 1016     & 0.1 &     &    &               &\phantom{$\times$} &          & &    &               \\
NGC 1023     & 0.0 &     &    & $\times$      &$\times$           &          & a & 0  & 1             \\
NGC 1316     & 4.1 & c   & 1  &               &$\times$           &          & E & 2  & 2,12              \\
NGC 1374     & 0.3 &     &    &               &\phantom{$\times$} & Offset   & a & 0  & 2             \\
NGC 1399     & 0.0 &     &    &               &$\times$           &          & a & 0.5& 2,4,8,12      \\
NGC 1426     & 0.0 &     &    &               &$\times$           &          & E & 1  & 9             \\
NGC 1427     & 0.0 &     &    & $\times$      &$\times$           &          & a & 0  & 2             \\
NGC 1439     & 2.4 & nr  & 2  &               &$\times$           &          & a & 0  & 14            \\
NGC 1700     & 0.2 & c   & 2  &               &\phantom{$\times$} &          & E & 1  & 3,4,9         \\
NGC 2300     & 0.0 &     &    &               &\phantom{$\times$} &          & a & 0  & 1,3,4         \\
NGC 2434     & 3.0 & nr  & 2  &               &$\times$           &          & a:     & 0: & 2,12              \\
NGC 2768     & \dots & s & 1  &               &\phantom{$\times$} &          & L2     & 2  & 1,7,9             \\
NGC 2778     & 0.3 & nr: & 2  &               &$\times$           &          & E:     & 2: & 3             \\
NGC 2974     & 2.5 &  s  & 1  &               &\phantom{$\times$} &          & E & 2  & 4,6,7,8,11,12   \\
NGC 3115     & 0.1 &     &    & $\times$      &$\times$           &          & a & 0  & 1,5            \\
NGC 3377     & 0.0 &  c  & 1  & $\times$      &\phantom{$\times$} &          & a &0.5 & 1,3,4,9        \\
NGC 3379     & 1.9 & nr  & 2  &               &\phantom{$\times$} &          & L2/T2::& 1  & 1,3,4,5           \\
NGC 3384     & 0.0 &     &    & $\times$      &$\times$           &          & a & 0  & 1             \\
NGC 3557     & \dots &nr & 2  &               &\phantom{$\times$} &          & E & 2  & 2,4,8,9      \\
NGC 3585     & 1.6 & nr  & 2  & $\times$      &\phantom{$\times$} &          & a & 0  & 6,8           \\
NGC 3607     & 8.7 & t   & 1  &               &\phantom{$\times$} &          & L 2     & 2  & 1,5,6         \\
NGC 3608     & 0.2 & nr: & 2  &               &\phantom{$\times$} &          & L2/S2: & 1  & 1,3,4             \\
NGC 3610     & 0.0 &     &    & $\times$      &\phantom{$\times$} &          & a & 0  & 1,4,9           \\
NGC 3640     & 0.2 &     &    &               &\phantom{$\times$} &          & a & 0  & 1,4           \\
NGC 3706     & 0.0 &     &    & $\times$      &\phantom{$\times$} & Bar      & a & 0  & 4,10             \\
NGC 3842     & 0.0 &     &    &               &\phantom{$\times$} &          & &    &               \\
NGC 3945     & 1.2 & c   & 1  &               &$\times$           &          & L2     & 2  & 1             \\
NGC 4026     & 0.0 & c   & 1  & $\times$      &$\times$           &          & a & 0  & 1             \\
NGC 4073     & 0.0 &     &    &               &\phantom{$\times$} & Min      & a & 0  & 14            \\
NGC 4125     & \dots & c & 1  &               &\phantom{$\times$} &          & T2     & 2  & 1,4,7         \\
NGC 4278     & 3.8 & s   & 1  &               &$\times$           &          & L2     & 3  & 1,3,4        \\
NGC 4291     & 0.0 &     &    &               &\phantom{$\times$} & Offset:  & a & 0  & 1             \\
NGC 4365     & 0.0 &     &    &               &$\times$           &          & a & 0  & 1,4,6,8        \\
NGC 4382     & 0.0 &     &    &               &\phantom{$\times$} & Min      & a & 0  & 1,6,8         \\
NGC 4406     & 0.0 &     &    &               &$\times$           & Min      & a & 0  & 1,6,8         \\
NGC 4458     & 0.2 &     &    & $\times$      &\phantom{$\times$} &          & &    &               \\
NGC 4472     & 0.7 & c   & 2  &               &$\times$           &          & S2::   & 1  & 1,3,5,6        \\
NGC 4473     & 0.0 &     &    & $\times$      &\phantom{$\times$} &          & a & 0  & 1,4,5,6           \\
NGC 4478     & 0.0 &     &    & $\times$      &$\times$           &          & a & 0  & 1,3               \\
NGC 4486B    & 0.0 &     &    &               &\phantom{$\times$} & Min      & &    &               \\
NGC 4494     & 5.3 & nr  & 2  &               &\phantom{$\times$} &          & L2::   & 1  & 1              \\
NGC 4552     & 0.5 & c   & 1  &               &$\times$           &          & T2:    & 1.5& 1,3,5,6,9      \\
NGC 4589     & 6.7 & c   & 1  &               &\phantom{$\times$} &          & L2     & 2  & 1,4,7             \\
NGC 4621     & 0.0 &     &    & $\times$      &\phantom{$\times$} &          & a & 0  & 1,4,6,8           \\
NGC 4649     & 0.0 &     &    &               &\phantom{$\times$} &          & a & 0  & 1,3,6,7,8         \\
NGC 4660     & 0.0 &     &    & $\times$      &\phantom{$\times$} &          & a & 0  & 1,4               \\
NGC 4709     & 0.0 &     &    &               &$\times$           &          & &    &               \\
NGC 4786     & \dots &nr & 2  &               &\phantom{$\times$} &          & E & 1  & 9             \\
NGC 4936     & \dots &c  & 1  &               &\phantom{$\times$} &          & E & 2  & 5,10             \\
NGC 5018     & \dots &t  & 1  &               &\phantom{$\times$} &          & E:     & 1: & 4,7,8,9,10     \\
NGC 5061     & 0.0 &     &    &               &\phantom{$\times$} &          & a & 0  & 4,8               \\
NGC 5322     & \dots &nr & 2  &               &\phantom{$\times$} &          & L2::   & 1  & 1,4,7,9        \\
NGC 5419     & 0.1 &     &    &               &$\times$ double?   &          & E:     & 1: & 2,5,12        \\
NGC 5557     & 0.1 &     &    &               &\phantom{$\times$} &          & a & 0  & 1,9             \\
NGC 5576     & 0.0 &     &    &               &\phantom{$\times$} & Offset:  & a & 0  & 1,4,9           \\
NGC 5813     & 0.5 & t   & 1  &               &\phantom{$\times$} &          & L2:    & 1  & 1,3,4,5,8         \\
NGC 5845     & \dots &nr & 2  &               &\phantom{$\times$} &          & &    &               \\
NGC 5982     & 0.1 &     &    &               &\phantom{$\times$} &          & L2::   & 1  & 1,9               \\
NGC 6776     & \dots &t  &    &               &\phantom{$\times$} &          & E & 2  & 2,5,9,10          \\
NGC 6849     & 0.0 &     &    &               &\phantom{$\times$} &          & a & 0  & 2,10                   \\
NGC 6876     & 0.0 & c   & 0  &               &\phantom{$\times$} & Min      & E & 1  & 2,5,6         \\
NGC 7213     & 4.0 & s   & 1  &               &$\times$           &          & S1     & 2  & 2,8               \\
NGC 7457     & 0.0 &     &    &               &$\times$           &          & c+a    & 0  & 1             \\
NGC 7619     & 0.0 &     &    &               &\phantom{$\times$} & Offset   & a & 0  & 1,3,5,6               \\
NGC 7626     & \dots &nr & 2  &               &\phantom{$\times$} &          & L2::   & 1  & 1,3,5,6,13      \\
NGC 7727     & 6.2 & s   & 1  &               &\phantom{$\times$} &          & E & 2  & 8               \\
NGC 7785     & 2.7 & c   & 2  &               &\phantom{$\times$} &          & E & 1  & 3,9                     \\
IC 1459      & 3.7 & t   & 1  &               &$\times$           &          & E & 3  & 4,5,6,9,11    \\
IC 3370      & \dots &s  & 1  &               &\phantom{$\times$} &          & E & 2  & 4,8                \\
IC 4296      & \dots &nr & 2  &               &\phantom{$\times$} &          & E & 1  & 2,5,6,9         \\
IC 4329      & 0.2 &     &    &               &\phantom{$\times$} &          & a & 0  & 10              \\
ESO 0462-015 & 0.1 &     &    &               &$\times$           &          & &    &                \\
\enddata
\label{tab:morph}
\tablecomments{Colons indicate weak and/or uncertain quantities.
Numbers under dust strength are hundredths of magnitudes
of dust absorption within $1''$ of the galaxy centers; ``\dots" indicates
dusty galaxies for which a Nuker law model could not be constructed;
the value for NGC~7457 was set to 0.0 by inspection.
Dust morphologies: nr = well defined
nuclear dust ring or disk, c = chaotic, disorganized dust patches, s = spiral
dust lanes, t = transition case between nuclear ring and non-ring
(i.e., between nr and c/s).  Dust concentration classes:
0 = dust present only outside $4''$ postage stamp in Figure \ref{fig:images},
1 = dust present both within $4''$ postage stamp and outside it,
2 = virtually all dust inside $4''$ postage stamp.  A ``$\times$" in the
stellar disk column indicates galaxies with stellar disks visible into
their centers in the subtracted-model postage stamps in Figure \ref{fig:images}.
The nucleus column indicates that a nucleus is detected
using the method explained in the text.  Peculiarities: ``Min" indicates a
brightness depression in the central pixel(s); ``Offset" indicates that the
central stellar peak is offset from the outer isophotes; ``Bar" indicates a
bright central bar (or edge-on disk).  Spectral types are taken preferentially
from Ho et al.~(1997), otherwise from the literature.  An
``a" indicates an absorption spectrum with no (or very small)
emission, ``L" means a LINER nucleus, ``T" a LINER/H II transition object,
``S" a Seyfert nucleus, and ``E" indicates
general emission of an unclassified type.  The ``c+a" for NGC~7457 indicates
a suspected nuclear continuum source plus absorption.
The numbers ``1" and ``2" under spectral type signify broad and narrow
permitted lines, respectively.  Emission-strength class:
0 = no emission detected, 1 = weak emission, 2 = strong emission,
3 = very strong emission.  Emission classes were determined by intercomparison
of data from the references in the last column.
Emission references: 1 = \cite{hfs97}; 2 = \cite{p86};
3 = \cite{g93}; 4 = \cite{goud94a}; 5 = \cite{macchetto96};
6 = \cite{shields}; 7 = \cite{kim}; 8 = \cite{vv88};
9 = \cite{c84}; 10 = \cite{carter}; 11 = \cite{buson};
12 = \cite{beuing}; 13 = \cite{vk99};
14 = \cite{humason}.}
\end{deluxetable}

\begin{deluxetable}{lcrrcrcrrl}
\tablecolumns{10}
\tablewidth{0pt}
\tablecaption{Central Structural Parameters}
\tablehead{\colhead{Galaxy}&\colhead{P}&\colhead{$r_b$ (pc)}
&\colhead{$\theta_b$}&\colhead{$I_b($V$)$}&\colhead{$\alpha$}
&\colhead{$\beta$}&\colhead{$\gamma$}&\colhead{$\gamma'$}&\colhead{Notes}}
\startdata
NGC 0507	&$\cap$&343.3&1.11&17.16& 1.26&1.75&$ 0.00$&$ 0.01$& \\
NGC 0584	&$\cap$& 47.1&0.44&15.22& 0.47&1.61&$-0.01$&$ 0.30$& \\
NGC 0596	&$\backslash$&  4.3&0.54&15.90& 0.45&1.59&$ 0.16$&$ 0.54$& \\
NGC 0741	&$\cap$&391.9&1.14&17.67& 2.27&1.29&$ 0.10$&$ 0.11$& \\
NGC 0821	&$\wedge$& 59.3&0.48&15.69& 0.43&1.71&$ 0.10$&$ 0.42$&$\backslash$ in Rav. \\
NGC 1016	&$\cap$&618.2&1.68&18.01& 0.98&1.90&$ 0.09$&$ 0.11$& \\
NGC 1023	&$\backslash$&  2.3&1.23&15.71& 7.23&1.15&$ 0.74$&$ 0.74$& \\
NGC 1316	&$\cap$& 41.8&0.38&14.41& 1.01&1.23&$-0.10$&$ 0.13$&$V-I=1.40$\\
NGC 1374	&$\cap$&  9.1&0.09&14.57& 1.87&1.08&$-0.09$&$-0.03$&$V-R=0.60$\\
NGC 1399	&$\cap$&324.3&3.17&17.23& 1.58&1.63&$ 0.09$&$ 0.09$&$V-R=0.60$\\
NGC 1426	&$\backslash$&  5.1&1.25&16.98& 0.51&1.80&$ 0.25$&$ 0.56$& \\
NGC 1427	&$\backslash$&  4.1&0.86&16.25& 0.79&1.67&$ 0.30$&$ 0.51$& \\
NGC 1439	&$\backslash$&  5.1&0.65&16.21& 4.82&1.48&$ 0.74$&$ 0.74$& \\
NGC 1700	&$\cap$& 11.8&0.06&13.58& 1.03&1.19&$-0.10$&$ 0.07$&$\backslash$ in L95 \\
NGC 2300	&$\cap$&330.1&2.24&17.53& 1.20&1.80&$ 0.07$&$ 0.08$& \\
NGC 2434	&$\backslash$&  4.4&5.41&19.10& 1.95&2.05&$ 0.75$&$ 0.75$& \\
NGC 2778	&$\backslash$&  4.7&0.35&16.06& 0.41&1.75&$ 0.33$&$ 0.83$& \\
NGC 2974	&$\backslash$&  4.4&0.44&15.41&15.72&1.10&$ 0.62$&$ 0.62$& \\
NGC 3115	&$\backslash$&  2.0&0.86&14.48& 4.30&1.09&$ 0.52$&$ 0.52$& \\
NGC 3377	&$\backslash$&  2.3&0.31&14.12& 0.30&1.96&$ 0.03$&$ 0.62$& \\
NGC 3379	&$\cap$&112.3&1.98&16.15& 1.54&1.54&$ 0.18$&$ 0.18$& \\
NGC 3384	&$\backslash$&  2.3&3.15&16.44&15.32&1.81&$ 0.71$&$ 0.71$& \\
NGC 3585	&$\wedge$& 37.0&0.36&14.72& 1.62&1.06&$ 0.31$&$ 0.31$& \\
NGC 3607	&$\cap$&128.4&2.43&16.87& 2.06&1.70&$ 0.26$&$ 0.26$& \\
NGC 3608	&$\cap$& 53.5&0.48&15.73& 0.92&1.50&$ 0.09$&$ 0.17$& \\
NGC 3610	&$\backslash$&  4.4&2.84&16.39&48.48&1.86&$ 0.76$&$ 0.76$& \\
NGC 3640	&$\cap$& 46.6&0.34&15.66& 0.64&1.30&$-0.10$&$ 0.03$&$\wedge$ in Rest  \\
NGC 3706	&$\cap$& 40.9&0.18&14.18&14.66&1.24&$-0.01$&$-0.01$& \\
NGC 3842	&$\cap$&507.9&1.08&17.76& 1.84&1.42&$ 0.15$&$ 0.15$& \\
NGC 3945	&$\backslash$&  3.9&7.38&18.62& 0.30&2.56&$-0.06$&$ 0.57$& \\
NGC 4026	&$\backslash$&  3.0&0.63&15.23& 0.39&1.78&$ 0.15$&$ 0.65$& \\
NGC 4073	&$\cap$&129.6&0.29&16.53& 4.43&0.99&$-0.08$&$-0.08$& \\
NGC 4278	&$\cap$&102.0&1.26&16.20& 1.40&1.46&$ 0.06$&$ 0.10$& \\
NGC 4291	&$\cap$& 72.7&0.60&15.66& 1.52&1.60&$ 0.01$&$ 0.02$& \\
NGC 4365	&$\cap$&231.4&2.21&16.88& 1.74&1.52&$ 0.07$&$ 0.09$& \\
NGC 4382	&$\cap$& 80.7&0.93&15.67& 1.13&1.39&$ 0.00$&$ 0.01$& \\
NGC 4406	&$\cap$& 79.8&0.92&16.03& 3.87&1.04&$-0.04$&$-0.04$& \\
NGC 4458	&$\cap$&  7.8&0.09&13.72& 4.55&1.40&$ 0.16$&$ 0.17$&$\backslash$ in L95 \\
NGC 4472	&$\cap$&211.7&2.44&16.63& 1.89&1.17&$ 0.01$&$ 0.01$& \\
NGC 4473	&$\cap$&386.2&4.45&17.24& 0.66&2.60&$-0.07$&$ 0.01$& \\
NGC 4478	&$\cap$& 19.1&0.22&15.46& 1.88&1.01&$-0.10$&$ 0.10$&$\backslash$ in L95 \\
NGC 4486B	&$\cap$& 13.9&0.16&14.53& 2.82&1.39&$-0.10$&$-0.10$& \\
NGC 4494	&$\backslash$&  3.5&2.82&17.19& 0.68&1.88&$ 0.52$&$ 0.55$& \\
NGC 4552	&$\cap$& 42.5&0.49&15.20& 1.97&1.17&$-0.10$&$-0.02$& \\
NGC 4589	&$\cap$& 67.9&0.56&16.08& 1.05&1.32&$ 0.21$&$ 0.25$& \\
NGC 4621	&$\backslash$&  3.5&1.00&15.76& 0.41&1.34&$ 0.75$&$ 0.85$& \\
NGC 4649	&$\cap$&437.4&5.04&17.28& 1.67&1.58&$ 0.16$&$ 0.16$& \\
NGC 4660	&$\backslash$&  3.5&1.76&16.34& 5.61&1.50&$ 0.91$&$ 0.91$& \\
NGC 4709	&$\cap$&222.4&1.24&17.46& 1.89&1.47&$ 0.27$&$ 0.28$& \\
NGC 5061	&$\cap$& 37.8&0.29&14.35& 1.64&1.43&$ 0.04$&$ 0.05$& \\
NGC 5419	&$\cap$&722.3&2.38&17.87& 1.43&1.69&$-0.10$&$ 0.03$& \\
NGC 5557	&$\cap$&204.4&0.80&16.40& 0.80&1.68&$ 0.02$&$ 0.07$&$\wedge$ in Rest \\
NGC 5576	&$\cap$&549.2&4.18&17.81& 0.43&2.73&$ 0.01$&$ 0.26$&$\backslash$ in Rest \\
NGC 5813	&$\cap$&162.0&1.17&16.86& 1.06&1.81&$-0.10$&$-0.08$& \\
NGC 5982	&$\cap$& 68.7&0.35&15.66& 2.07&1.03&$ 0.05$&$ 0.05$& \\
NGC 6849	&$\cap$&138.5&0.34&17.03& 1.01&1.21&$ 0.01$&$ 0.08$& \\
NGC 6876	&$\cap$&139.6&0.50&16.99&12.07&0.75&$ 0.00$&$ 0.00$& \\
NGC 7213	&$\cap$&383.0&3.48&17.71& 1.03&3.10&$ 0.06$&$ 0.21$& \\
NGC 7457	&$\backslash$&  2.7&0.22&16.33& 1.01&1.05&$-0.10$&$ 0.61$&$\wedge$ in Rav. \\
NGC 7619	&$\cap$&220.3&0.81&16.40& 1.08&1.62&$-0.02$&$ 0.01$& \\
NGC 7727	&$\wedge$&601.1&5.74&18.42& 0.57&1.93&$ 0.43$&$ 0.49$& \\
NGC 7785	&$\cap$& 14.5&0.06&15.09& 0.95&0.92&$-0.10$&$ 0.05$& \\
IC 1459	&$\cap$&293.6&1.96&16.36& 0.80&2.08&$-0.10$&$ 0.15$& \\
IC 4329	&$\cap$&261.2&0.79&16.37& 2.87&1.31&$ 0.01$&$ 0.01$& \\
ESO 462-015	&$\backslash$& 15.6&0.41&16.59& 0.20&1.77&$-0.10$&$ 0.56$& \\
\enddata
\label{tab:cen}
\tablecomments{The profile type, P, is $\backslash=$ power-law,
$\wedge=$ intermediate form, and $\cap=$ core. $\theta_b$ is the
angular break radius in arcseconds along the major axis.
The Nuker law parameters are defined in equation (\ref{eqn:nuker}).
The surface brightness, $I_b,$ is in units of $V$ mags per square arcsec.
No Galactic absorption corrections have been applied, but NGC 1316, 1374,
and 1399 have been transformed to the $V$ band using the color terms in
the notes column.
$\gamma'$ is the power-law slope $-d\log I/d\log r$ at the inner resolution
limit, which is $0\asec02$ for dithered images, or $0\asec04$ otherwise.
The notes column lists references
with differing profile classifications, where L95=\citet{l95},
Rest=\citet{rest}, and Rav.=\citet{rav}.}
\end{deluxetable}

\begin{deluxetable}{lccc}
\tablecolumns{4}
\tablewidth{0pt}
\tablecaption{Average Ellipticity}
\tablehead{\colhead{ }&\colhead{ }&\colhead{Inner}&\colhead{Outer}\\
\colhead{Galaxy}&\colhead{P}&\colhead{Ellipticity}&\colhead{Ellipticity}}
\startdata
NGC  507&$\cap$&0.229&0.190\\
NGC  584&$\cap$&0.346&0.255\\
NGC  596&$\backslash$&0.123&0.072\\
NGC  741&$\cap$&0.043&0.147\\
NGC  821&$\wedge$&0.370&0.382\\
NGC 1016&$\cap$&0.099&0.071\\
NGC 1023&$\backslash$&0.351&0.201\\
NGC 1316&$\cap$&0.242&0.367\\
NGC 1374&$\cap$&0.036&0.110\\
NGC 1399&$\cap$&0.078&0.125\\
NGC 1426&$\backslash$&0.320&0.315\\
NGC 1427&$\backslash$&0.287&0.293\\
NGC 1439&$\backslash$&0.274&0.148\\
NGC 1700&$\cap$&0.318&0.294\\
NGC 2300&$\cap$&0.196&0.232\\
NGC 2434&$\backslash$&0.083&0.067\\
NGC 2778&$\backslash$&0.111&0.204\\
NGC 2974&$\backslash$&0.296&0.301\\
NGC 3115&$\backslash$&0.537&0.548\\
NGC 3377&$\backslash$&0.360&0.509\\
NGC 3379&$\cap$&0.102&0.097\\
NGC 3384&$\backslash$&0.415&0.180\\
NGC 3585&$\wedge$&0.283&0.456\\
NGC 3607&$\cap$&0.129&0.159\\
NGC 3608&$\cap$&0.108&0.184\\
NGC 3610&$\backslash$&0.437&0.516\\
NGC 3640&$\cap$&0.110&0.212\\
NGC 3842&$\cap$&0.044&0.174\\
NGC 3945&$\backslash$&0.191&0.299\\
NGC 4026&$\backslash$&0.469&0.385\\
NGC 4278&$\cap$&0.155&0.167\\
NGC 4291&$\cap$&0.260&0.247\\
NGC 4365&$\cap$&0.286&0.248\\
NGC 4382&$\cap$&0.262&0.240\\
NGC 4458&$\cap$&0.573&0.187\\
NGC 4472&$\cap$&0.088&0.110\\
NGC 4473&$\cap$&0.469&0.412\\
NGC 4478&$\cap$&0.311&0.220\\
NGC 4486&$\cap$&0.431&0.124\\
NGC 4494&$\backslash$&0.225&0.153\\
NGC 4552&$\cap$&0.138&0.050\\
NGC 4589&$\cap$&0.399&0.197\\
NGC 4621&$\backslash$&0.299&0.351\\
NGC 4649&$\cap$&0.072&0.140\\
NGC 4660&$\backslash$&0.396&0.404\\
NGC 4709&$\cap$&0.233&0.184\\
NGC 5061&$\cap$&0.060&0.038\\
NGC 5419&$\cap$&0.180&0.208\\
NGC 5557&$\cap$&0.089&0.209\\
NGC 5576&$\cap$&0.242&0.302\\
NGC 5813&$\cap$&0.124&0.107\\
NGC 5982&$\cap$&0.031&0.273\\
NGC 6849&$\cap$&0.056&0.353\\
NGC 7213&$\cap$&0.108&0.110\\
NGC 7457&$\backslash$&0.064&0.371\\
NGC 7619&$\cap$&0.392&0.266\\
NGC 7727&$\wedge$&0.262&0.243\\
NGC 7785&$\cap$&0.266&0.367\\
IC 1459&$\cap$&0.260&0.261\\
IC 4329&$\cap$&0.168&0.152\\
ESO 462-15&$\backslash$&0.170&0.271\\
\enddata
\label{tab:ellip}
\tablecomments{Luminosity-weighted average isophote ellipticity
for $r\leq r_b$ and $r>r_b.$
Parameters were not measured for NGC 4073, 4406, 4486B, and 6876,
which have central surface brightness minima.}
\end{deluxetable}

\begin{deluxetable}{lcccrrl}
\tablecolumns{7}
\tablewidth{0pt}
\tablecaption{Nuclei Luminosities and Colors}
\tablehead{\colhead{Galaxy}&\colhead{P}&\colhead{$m_V$}&\colhead{$V-I$}
&\colhead{$\Delta(V-I)$}&\colhead{$M_V$}&\colhead{$L/L_\odot(V)$}}
\startdata
NGC 0596&$\backslash$&   21.8&   \dots&   \dots& $-$10.0&$9\times10^{5}$ \\
NGC 0741&$\cap$&   21.9&   \dots&   \dots& $-$12.5&$9\times10^{6}$ \\
NGC 1023&$\backslash$&   19.5&   0.9&  $-$0.4& $-$11.1&$3\times10^{6}$ \\
NGC 1316&$\cap$&   19.9&   \dots&   \dots& $-$12.0&$5\times10^{6}$ \\
NGC 1399&$\cap$&   24.4&   \dots&   \dots&  $-$7.2&$7\times10^{4}$ \\
NGC 1426&$\backslash$&   19.6&   \dots&   \dots& $-$12.5&$9\times10^{6}$ \\
NGC 1427&$\backslash$&   19.6&   1.3&   0.0& $-$12.0&$6\times10^{6}$ \\
NGC 1439&$\backslash$&   20.2&   1.1&  $-$0.2& $-$11.9&$5\times10^{6}$ \\
NGC 2434&$\backslash$&   20.3&   0.7&  $-$0.5& $-$12.3&$7\times10^{6}$ \\
NGC 2778&$\backslash$&   20.0&   1.3&  $-$0.1& $-$11.9&$5\times10^{6}$ \\
NGC 3115&$\backslash$&   17.8&   1.2&  $-$0.1& $-$12.3&$8\times10^{6}$ \\
NGC 3384&$\backslash$&   18.9&   1.3&   0.0& $-$11.5&$4\times10^{6}$ \\
NGC 3945&$\backslash$&   20.9&   1.5&   0.2& $-$10.7&$2\times10^{6}$ \\
NGC 4026&$\backslash$&   19.5&   1.3&   0.0& $-$11.6&$4\times10^{6}$ \\
NGC 4278&$\cap$&   19.5&   0.4&  $-$0.8& $-$11.7&$4\times10^{6}$ \\
NGC 4365&$\cap$&   21.0&   0.6&  $-$0.7& $-$10.6&$2\times10^{6}$ \\
NGC 4406&$\cap$&   22.9&   1.2&   0.0&  $-$8.4&$2\times10^{5}$ \\
NGC 4472&$\cap$&   22.7&   1.0&  $-$0.4&  $-$8.6&$2\times10^{5}$ \\
NGC 4478&$\cap$&   19.7&   \dots&   \dots& $-$11.6&$4\times10^{6}$ \\
NGC 4552&$\cap$&   20.5&   1.3&   0.0& $-$10.9&$2\times10^{6}$ \\
NGC 4709&$\cap$&   23.2&   \dots&   \dots& $-$10.0&$8\times10^{5}$ \\
NGC 5419&$\cap$&   19.8&   \dots&   \dots& $-$14.4&$5\times10^{7}$ \\
NGC 7213&$\cap$&   16.6&   0.6&  $-$0.8& $-$15.2&$1\times10^{8}$ \\
NGC 7457&$\backslash$&   18.1&   1.2&   0.1& $-$12.7&$1\times10^{7}$ \\
 IC 1459&$\cap$&   18.4&   1.0&  $-$0.4& $-$14.0&$4\times10^{7}$ \\
\enddata
\label{tab:nuclei}
\tablecomments{All photometry has been corrected for Galactic extinction
$\Delta(V-I)$ gives the color of the nucleus relative to the host galaxy
immediately outside the nucleus.  $M_V$ is the nuclear absolute magnitude
estimated by subtracting off the Nuker law fit to the pixels beyond
the nucleus. The last column gives the $V$ band nuclear luminosity
in solar units.}
\end{deluxetable}

\begin{deluxetable}{lcc}
\tablecolumns{3}
\tablewidth{0pt}
\tablecaption{Central Color Gradients}
\tablehead{\colhead{Galaxy}&\colhead{$V-I$}
&\colhead{$d(V-I)/d\log(r)$}}
\startdata
NGC 0584&$1.266\pm0.003$&$-0.084\pm0.007$\\
NGC 0821&$1.350\pm0.001$&$-0.056\pm0.002$\\
NGC 1023&$1.258\pm0.002$&$-0.039\pm0.004$\\
NGC 1427&$1.237\pm0.002$&$-0.066\pm0.004$\\
NGC 1439&$1.274\pm0.004$&$-0.125\pm0.008$\\
NGC 1700&$1.289\pm0.002$&$-0.051\pm0.003$\\
NGC 2300&$1.329\pm0.001$&$+0.005\pm0.002$\\
NGC 2434&$1.231\pm0.004$&$-0.092\pm0.008$\\
NGC 2778&$1.280\pm0.002$&$-0.053\pm0.004$\\
NGC 3115&$1.278\pm0.002$&$-0.046\pm0.005$\\
NGC 3377&$1.229\pm0.003$&$-0.028\pm0.006$\\
NGC 3379&$1.282\pm0.001$&$-0.029\pm0.003$\\
NGC 3384&$1.252\pm0.002$&$-0.047\pm0.004$\\
NGC 3607&$1.360\pm0.013$&$-0.058\pm0.026$\\
NGC 3608&$1.289\pm0.002$&$-0.068\pm0.004$\\
NGC 3945&$1.278\pm0.002$&$-0.063\pm0.003$\\
NGC 4026&$1.258\pm0.003$&$-0.041\pm0.005$\\
NGC 4278&$1.264\pm0.003$&$-0.028\pm0.006$\\
NGC 4291&$1.271\pm0.001$&$-0.036\pm0.003$\\
NGC 4365&$1.325\pm0.002$&$-0.034\pm0.004$\\
NGC 4382&$1.108\pm0.001$&$+0.006\pm0.003$\\
NGC 4406&$1.253\pm0.003$&$-0.028\pm0.006$\\
NGC 4458&$1.176\pm0.002$&$-0.010\pm0.004$\\
NGC 4472&$1.330\pm0.002$&$-0.027\pm0.004$\\
NGC 4473&$1.280\pm0.001$&$-0.039\pm0.002$\\
NGC 4486B&$1.247\pm0.002$&$-0.016\pm0.003$\\
NGC 4494&$1.257\pm0.006$&$-0.137\pm0.013$\\
NGC 4552&$1.289\pm0.001$&$-0.027\pm0.002$\\
NGC 4589&$1.331\pm0.004$&$-0.142\pm0.010$\\
NGC 4621&$1.300\pm0.003$&$-0.018\pm0.005$\\
NGC 4649&$1.338\pm0.001$&$-0.008\pm0.002$\\
NGC 4660&$1.296\pm0.001$&$-0.051\pm0.003$\\
NGC 5576&$1.183\pm0.001$&$-0.022\pm0.003$\\
NGC 5813&$1.310\pm0.013$&$+0.018\pm0.026$\\
NGC 5982&$1.264\pm0.003$&$-0.033\pm0.006$\\
NGC 7213&$1.440\pm0.028$&$+0.127\pm0.057$\\
NGC 7457&$1.122\pm0.009$&$+0.027\pm0.019$\\
NGC 7619&$1.353\pm0.002$&$-0.059\pm0.005$\\
NGC 7727&$0.570\pm0.007$&$-0.122\pm0.015$\\
IC 1459&$1.369\pm0.004$&$-0.087\pm0.008$\\
\enddata
\label{tab:vmi}
\tablecomments{$V-I$ colors are given at $1''$ from the galaxy centers.
Colors have been corrected for Galactic extinction.}
\end{deluxetable}

\begin{deluxetable}{lccrrcc}
\tablecolumns{7}
\tablewidth{0pt}
\tablecaption{Isophote PA Gradients and Twists}
\tablehead{\colhead{ }&\colhead{ }&\colhead{ }
&\colhead{Inner}&\colhead{Outer}&\colhead{ }&\colhead{ } \\
\colhead{ }&\colhead{Twist Radius}
&\colhead{ }&\colhead{Gradient}&\colhead{Gradient}
&\colhead{Inner}&\colhead{Outer}\\
\colhead{Galaxy}&\colhead{(arcsec)}
&\colhead{$\gamma'$}&\colhead{${\Delta \rm{PA(deg)}\over\Delta\log(r)}$}&\colhead{${\Delta \rm{PA(deg)}\over\Delta\log(r)}$}&\colhead{Twist}&\colhead{Twist}}
\startdata
NGC 0507&0.50&0.47&$ -64.2$&$   4.2$&$1.1\times10^{-2}$&$1.0\times10^{-2}$\\
NGC 0584&0.67&0.87&$  -4.1$&$   2.1$&$6.4\times10^{-3}$&$4.6\times10^{-3}$\\
NGC 0596&3.82&1.17&$  -0.6$&$ -92.5$&$1.0\times10^{-2}$&$1.4\times10^{-2}$\\
NGC 0741&1.08&0.67&$ 116.5$&$  -4.9$&$8.3\times10^{-3}$&$3.2\times10^{-3}$\\
NGC 0821&0.78&0.99&$   3.9$&$   0.1$&$4.0\times10^{-3}$&$1.3\times10^{-3}$\\
NGC 1016&0.46&0.48&$ -70.3$&$  -1.5$&$8.2\times10^{-3}$&$1.9\times10^{-3}$\\
NGC 1023&1.44&1.05&$   1.0$&$   5.0$&$7.0\times10^{-3}$&$2.7\times10^{-3}$\\
NGC 1316&1.08&0.89&$  40.0$&$  -6.0$&$2.7\times10^{-2}$&$1.3\times10^{-2}$\\
NGC 1374&0.32&0.98&$  10.5$&$   0.1$&$2.1\times10^{-3}$&$1.1\times10^{-3}$\\
NGC 1399&0.48&0.16&$   7.4$&$   1.8$&$3.9\times10^{-3}$&$2.7\times10^{-3}$\\
NGC 1426&0.30&0.75&$   5.2$&$   3.4$&$1.3\times10^{-2}$&$7.9\times10^{-3}$\\
NGC 1427&0.55&0.86&$   8.6$&$  -0.4$&$5.1\times10^{-3}$&$2.6\times10^{-3}$\\
NGC 1439&0.78&1.27&$  -0.9$&$ -15.7$&$3.8\times10^{-2}$&$1.8\times10^{-2}$\\
NGC 1700&0.30&0.98&$   6.0$&$  -1.3$&$8.4\times10^{-3}$&$1.0\times10^{-2}$\\
NGC 2300&0.57&0.34&$  46.6$&$  -4.7$&$3.8\times10^{-3}$&$3.3\times10^{-3}$\\
NGC 2434&0.50&0.76&$-113.2$&$  22.9$&$2.1\times10^{-2}$&$5.0\times10^{-3}$\\
NGC 2778&4.68&1.39&$   6.0$&$ -11.7$&\dots&\dots\\
NGC 2974&0.39&0.68&$ -15.7$&$  -2.3$&$2.7\times10^{-2}$&$4.7\times10^{-2}$\\
NGC 3115&0.78&0.75&$   1.0$&$   0.0$&$3.4\times10^{-2}$&$4.6\times10^{-3}$\\
NGC 3377&1.76&1.24&$   0.4$&$   0.0$&$2.2\times10^{-3}$&$3.0\times10^{-3}$\\
NGC 3379&0.67&0.39&$ -50.0$&$   7.4$&$5.9\times10^{-3}$&$2.9\times10^{-3}$\\
NGC 3384&4.68&1.81&$   0.6$&$  19.4$&\dots&\dots\\
NGC 3585&0.43&0.74&$ -14.6$&$  -0.6$&$7.3\times10^{-3}$&$3.5\times10^{-3}$\\
NGC 3607&1.08&0.49&$ -87.8$&$  58.2$&$2.1\times10^{-2}$&$4.6\times10^{-2}$\\
NGC 3608&0.57&0.85&$  34.1$&$   1.0$&$6.9\times10^{-3}$&$7.2\times10^{-4}$\\
NGC 3610&1.04&0.76&$   9.5$&$  -6.6$&$1.3\times10^{-2}$&$1.0\times10^{-2}$\\
NGC 3640&0.34&0.60&$  15.8$&$ -11.9$&$6.6\times10^{-3}$&$6.2\times10^{-3}$\\
NGC 3842&0.78&0.60&$-148.0$&$  18.8$&$8.0\times10^{-3}$&$2.6\times10^{-3}$\\
NGC 3945&2.87&1.07&$ -23.8$&$  57.1$&$2.7\times10^{-2}$&$6.2\times10^{-2}$\\
NGC 4026&0.30&0.84&$  -2.3$&$   1.2$&$8.9\times10^{-3}$&$1.6\times10^{-2}$\\
NGC 4278&0.46&0.33&$  61.1$&$   4.7$&$1.0\times10^{-2}$&$6.4\times10^{-3}$\\
NGC 4291&3.38&1.50&$   0.8$&$  -2.0$&$1.6\times10^{-3}$&$2.7\times10^{-3}$\\
NGC 4365&1.76&0.65&$  -3.4$&$   0.6$&$1.0\times10^{-3}$&$1.3\times10^{-3}$\\
NGC 4382&1.50&0.88&$ -16.4$&$   0.0$&$6.5\times10^{-3}$&$5.0\times10^{-3}$\\
NGC 4458&1.50&1.40&$   0.0$&$   0.7$&$2.9\times10^{-3}$&$7.4\times10^{-4}$\\
NGC 4472&0.67&0.10&$ -56.3$&$ -12.3$&$3.7\times10^{-3}$&$2.0\times10^{-3}$\\
NGC 4473&0.46&0.42&$   6.4$&$   0.2$&$2.6\times10^{-3}$&$1.1\times10^{-3}$\\
NGC 4478&0.30&0.60&$  25.8$&$  -2.7$&$9.8\times10^{-3}$&$1.7\times10^{-2}$\\
NGC 4486B&4.68&1.39&$  -2.2$&$ -39.6$&\dots&\dots\\
NGC 4494&0.41&0.81&$  16.9$&$ -12.7$&$1.5\times10^{-2}$&$1.2\times10^{-2}$\\
NGC 4552&0.67&0.72&$ -51.2$&$   4.8$&$4.7\times10^{-3}$&$1.9\times10^{-3}$\\
NGC 4589&2.08&1.10&$  -9.4$&$  -8.4$&$1.6\times10^{-2}$&$3.8\times10^{-3}$\\
NGC 4621&0.32&0.98&$ -11.0$&$   0.1$&$7.3\times10^{-3}$&$7.6\times10^{-4}$\\
NGC 4649&2.44&0.49&$ -13.8$&$  11.7$&$9.1\times10^{-4}$&$1.5\times10^{-3}$\\
NGC 4660&4.68&1.50&$  -0.5$&$ -19.0$&\dots&\dots\\
NGC 4709&0.36&0.38&$ -21.3$&$   3.0$&$4.2\times10^{-3}$&$2.7\times10^{-3}$\\
NGC 5061&3.24&1.41&$  -6.2$&$-115.9$&$6.8\times10^{-3}$&$2.5\times10^{-2}$\\
NGC 5419&0.92&0.27&$  34.4$&$  -5.7$&$1.7\times10^{-2}$&$4.7\times10^{-3}$\\
NGC 5557&0.48&0.68&$  45.8$&$  -5.7$&$3.8\times10^{-3}$&$5.6\times10^{-3}$\\
NGC 5576&0.32&0.68&$  15.5$&$  -2.8$&$3.9\times10^{-3}$&$4.1\times10^{-3}$\\
NGC 5813&0.57&0.51&$ -64.0$&$  25.4$&$3.4\times10^{-3}$&$4.7\times10^{-3}$\\
NGC 5982&0.27&0.41&$-134.6$&$  14.5$&$4.3\times10^{-3}$&$4.3\times10^{-3}$\\
NGC 6849&0.41&0.67&$  81.8$&$   1.1$&$5.7\times10^{-3}$&$3.5\times10^{-3}$\\
NGC 7213&0.54&0.45&$  50.2$&$ -18.5$&$3.3\times10^{-2}$&$6.9\times10^{-2}$\\
NGC 7457&0.27&0.53&$  54.8$&$  -1.7$&\dots&\dots\\
NGC 7619&0.67&0.71&$  -3.3$&$   0.4$&$1.9\times10^{-3}$&$1.9\times10^{-3}$\\
NGC 7727&4.68&1.14&$  15.9$&$ -55.0$&\dots&\dots\\
NGC 7785&1.99&0.88&$   3.4$&$  -1.6$&$8.7\times10^{-3}$&$5.3\times10^{-3}$\\
IC 1459&0.50&0.45&$  41.4$&$   0.1$&$1.7\times10^{-2}$&$2.2\times10^{-3}$\\
IC 4329&0.57&0.37&$ -62.2$&$   3.0$&$3.3\times10^{-3}$&$2.1\times10^{-3}$\\
ESO 0462-015&0.36&0.83&$ -97.7$&$   2.0$&$2.7\times10^{-2}$&$3.1\times10^{-3}$\\
\enddata
\label{tab:bend}
\tablecomments{The twist values are measured over the domains of the
inner and outer gradients, using the expression in equation (\ref{eqn:twist}).
$\gamma'$ gives the local power-law slope
of the brightness profile at the twist radius. The outer domain was too
limited for twists to be measured for NGC 2778, 3384, 4660, 7457, and 7727.
Parameters were also not measured for NGC 4073, 4406, 4486B, and 6876,
which have central surface brightness minima.}
\end{deluxetable}
\begin{deluxetable}{lcccl}
\tablecolumns{5}
\tablewidth{0pt}
\tablecaption{Frequencies of Dust Morphologies}
\tablehead{\colhead{\hbox to 50pt{\hfill}}&
           \colhead{This paper}&
           \colhead{\citet{t01}}&
           \colhead{\citet{l95}}&
           \colhead{Total (\%)}}
\startdata
No dust         &  39  &  34  &  21  &  94\phantom{7} (53\%\ )  \\
Nuclear rings   &  14  &  16  &   1  &  31\phantom{7} (18\%\ )  \\
Non-rings       &  24  &  23  &   5  &  52\phantom{7} (29\%\ )  \\
Total           &  77  &  73  &  27  &  177 (100\%)  \\
\enddata
\label{tab:dustfreq}
\tablecomments{The frequency of dust morphologies in the three
papers defining the dust sample is given.
``Non-rings" include chaotics (c), spirals (s), and
transitional types (t).  The second column counts dust morphologies
for the present sample.  The \citet{t01} column only counts galaxies
in that paper not already counted by the present paper;
the \citet{l95} refers to additional
galaxies not in the present paper or in \citet{t01}.}
\end{deluxetable}

\begin{deluxetable}{lcccc}
\tablecolumns{5}
\tablewidth{0pt}
\tablecaption{Dust Strength versus Optical Emission Strength: Present Paper}
\tablehead{\colhead{Emission strength}&
           \colhead{0}&
           \colhead{1}&
           \colhead{2}&
           \colhead{3}}
\startdata
Dust Strength $>2.5$           &   0    &   4    &   11    &   2            \\
Dust Strength $=0.3-2.5$       &   2    &  5.5   &   2.5   &   0            \\
Dust strength $<0.3$           &   31   &   7    &   0     &   0            \\
\enddata
\label{tab:dustvsemiss}
\tablecomments{Dust strength vs.~optical emission-strength class from
Table \ref{tab:morph}.  Galaxies that are too dusty
for a reference model (Figure \ref{fig:obs_im})
have been assigned dust strengths $>$ 2.5
except for NGC~4786, 5845, 6776, and 7626, which were assigned
to the middle dust strength category.
Galaxies on bin borders have been assigned to both neighboring
bins with half weight.  Galaxies with colons on their emission strengths
in Table \ref{tab:morph} have not been used.}
\end{deluxetable}
\clearpage
\begin{figure}
\plotone{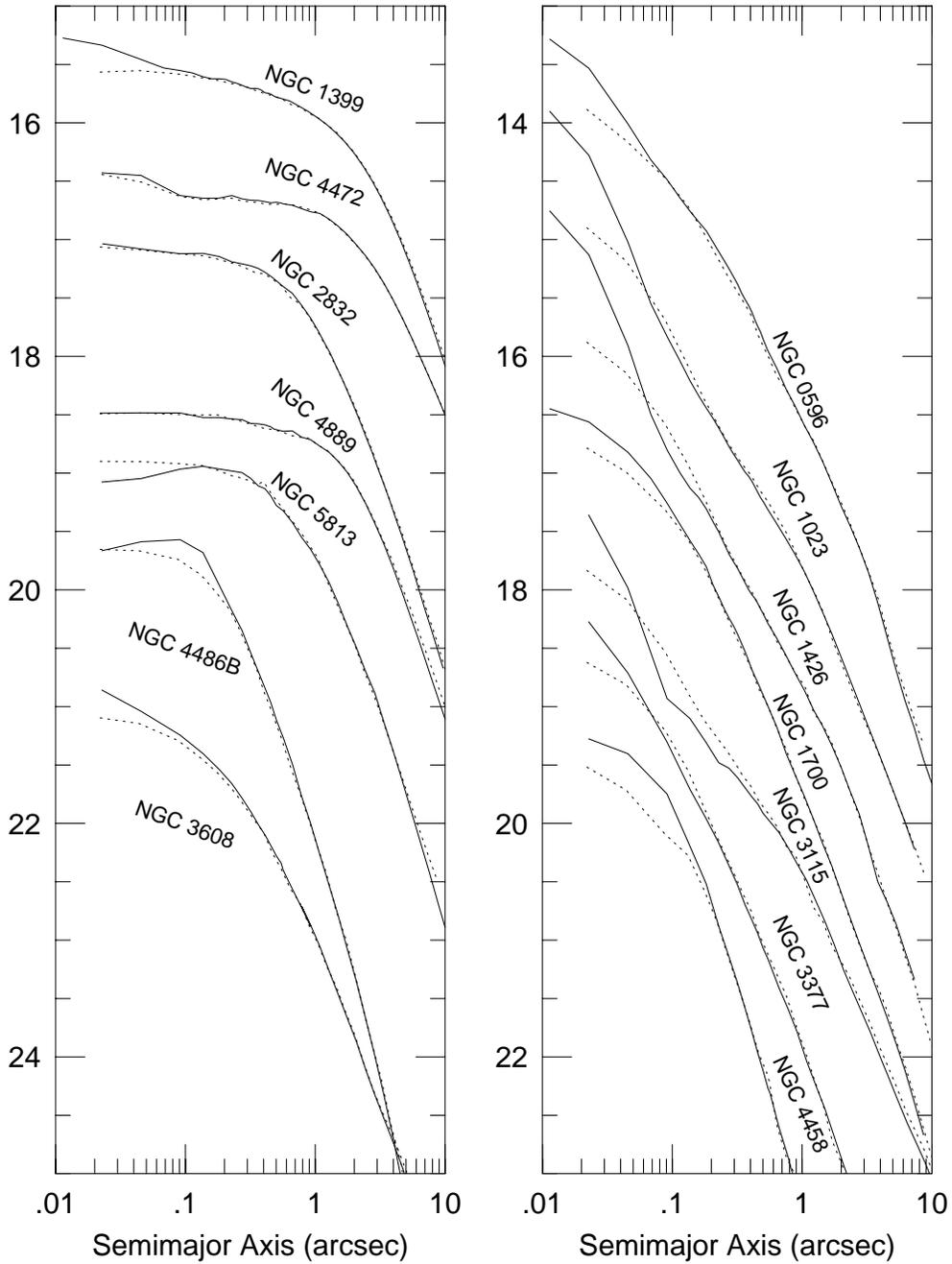}
\caption{Deconvolved WFPC2 brightness profiles (solid lines) are compared
to deconvolved WFPC1 profiles (dashed lines) for the same galaxies.
The vertical scale is in magnitudes per square arcsecond.
The WFPC2 profiles for NGC 2832 and NGC 4889 were published in \citet{laine}.
All other WFPC2 profiles can also be seen in Figure \ref{fig:surf}.
The WFPC1 and WFPC2 profiles agree extremely well for $r>0\asec1,$
except for some galaxies with strong nuclei.}
\label{fig:decon}
\end{figure}
\begin{figure}
\plotone{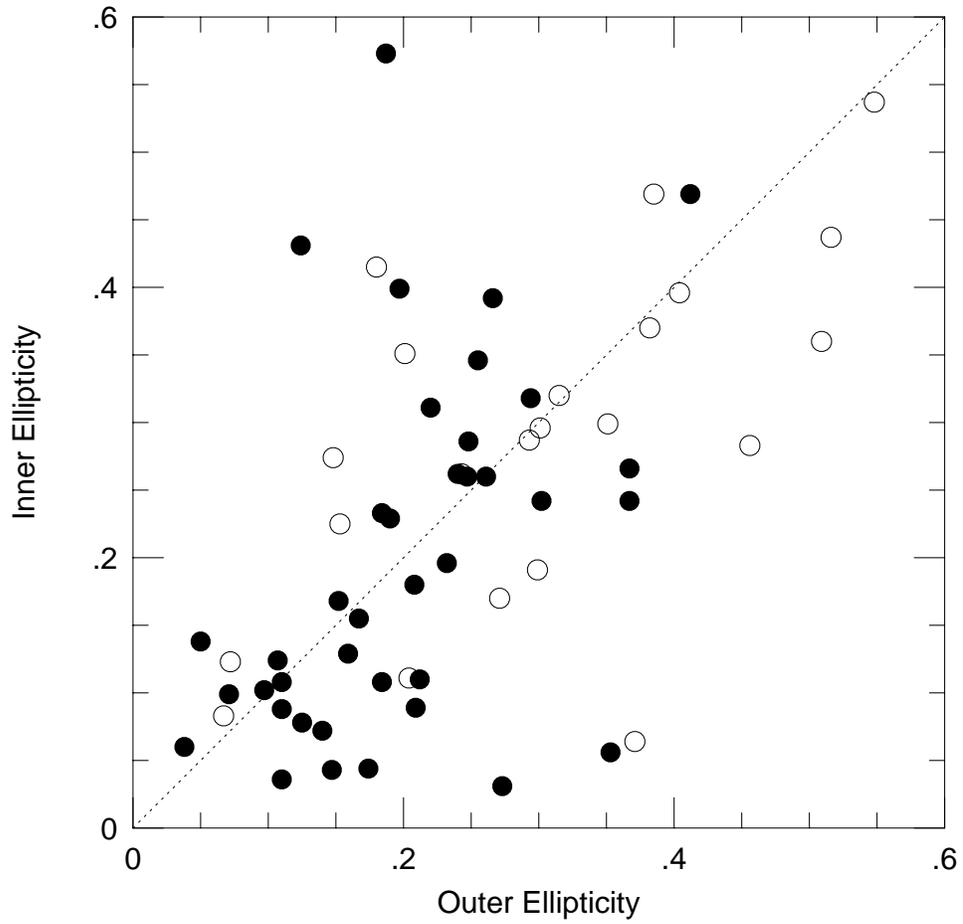}
\caption{Inner ($r\leq r_b$) luminosity-weighted isophote ellipticity is
plotted against outer isophote ellipticity.  Solid symbols are core galaxies;
open symbols are power-law or intermediate galaxies.
There is no strong average trend for isophote ellipticity to change with
radius for either type of galaxy.}
\label{fig:ellip}
\end{figure}
\begin{figure}
\plotone{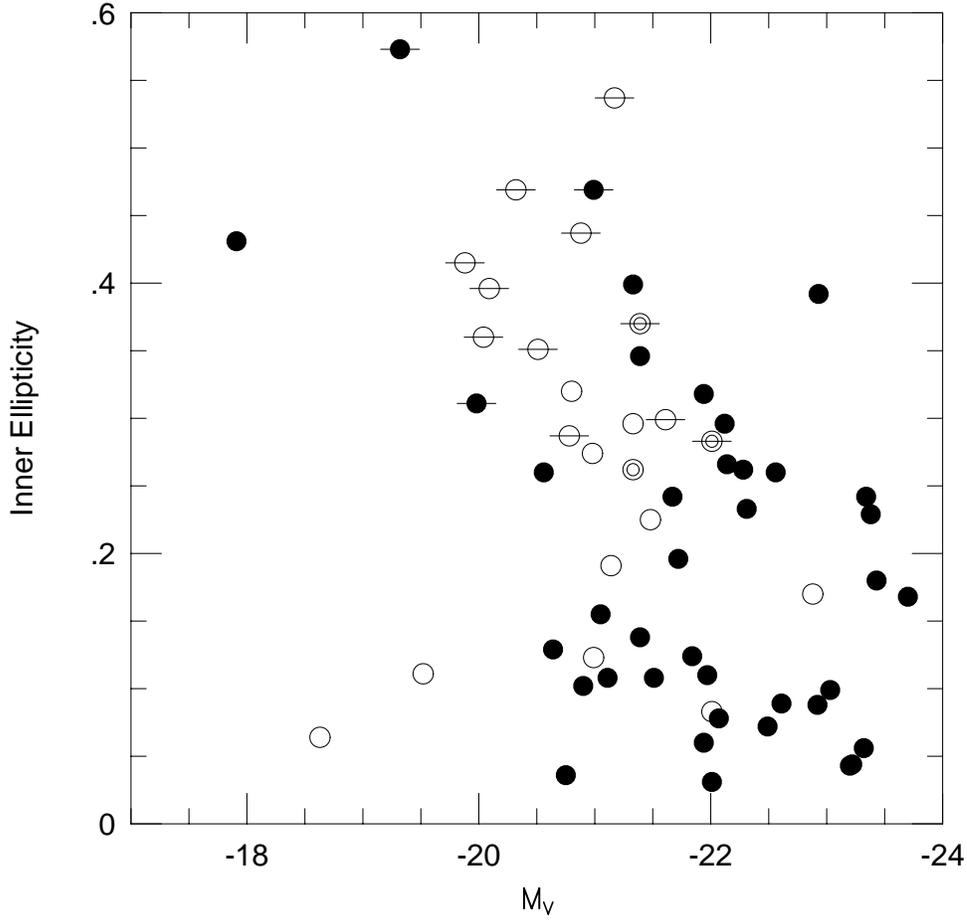}
\caption{Inner ($r\leq r_b$) luminosity-weighted isophote ellipticity is
plotted as a function of total galaxy luminosity.  Solid symbols are core
galaxies, open symbols are power-law galaxies, and intermediate galaxies
are plotted as double open circles.  Galaxies with {\it inner} stellar
disks are indicated with horizontal lines.  Power-law galaxies tend
to have higher inner ellipticity than core galaxies.  Nearly all power-law
galaxies with $\epsilon\geq0.3$ have inner disks, implying that they
are also present in the rounder power-law galaxies, but are not seen due to
unfavorable viewing angles.  Disks are visible in flattened core
and intermediate galaxies fainter than $M_V\approx-21,$ suggesting
that these are transitional objects.  Disks are not seen in bright
core galaxies.}
\label{fig:ellip_mag}
\end{figure}
\begin{figure}
\plotone{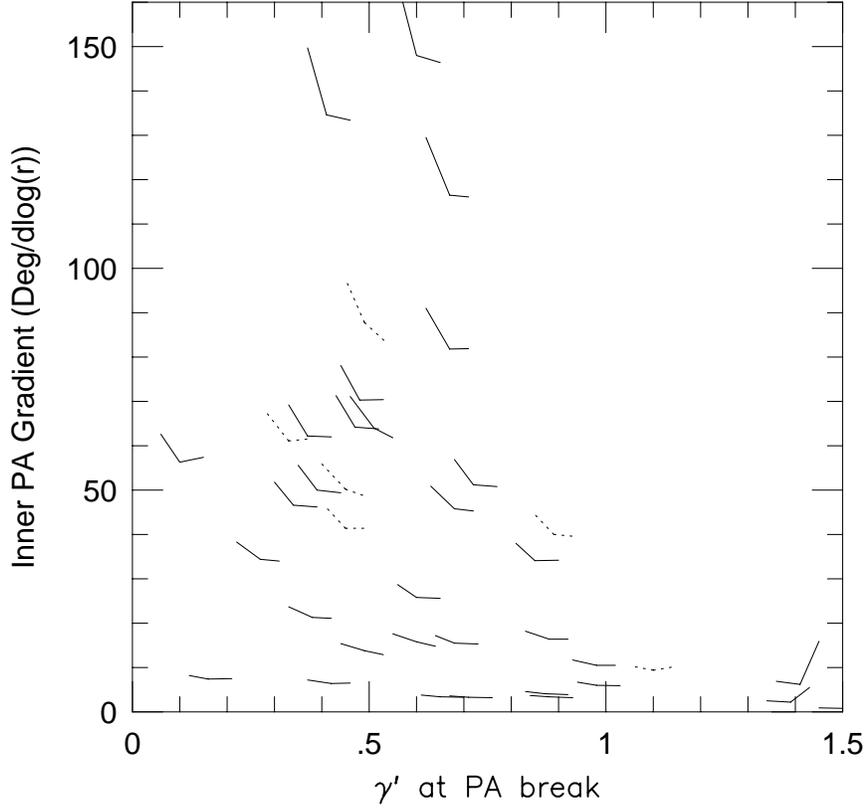}
\caption{Absolute value of the inner gradients in isophote position
angle for core galaxies only are plotted as a function of the
local logarithmic surface brightness slope, $\gamma',$ at the
radius dividing the inner and outer gradients.  The wedge symbols
also encode the value of the inner gradients in the slope of their
left line segments and the outer gradients in the right segments.
Dotted symbols are those galaxies that may be strongly affected by dust.
Radii of the break or ``hinge'' in the PA gradients cluster near
$\gamma'=0.5,$ which lies well within the break radii.}
\label{fig:bend}
\end{figure}
\begin{figure}
\plotone{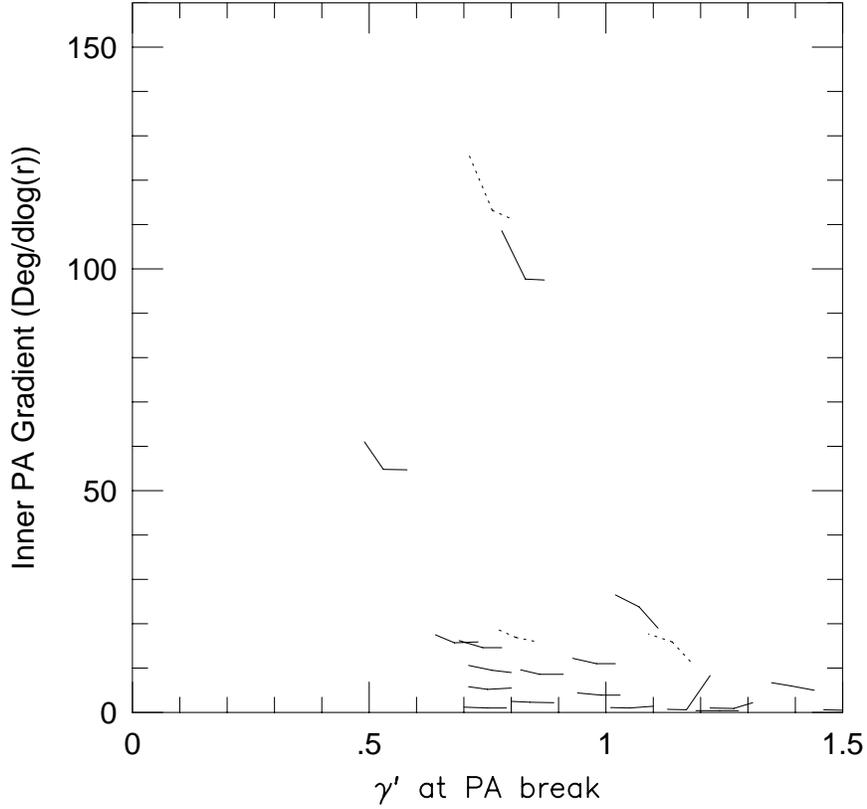}
\caption{Absolute value of the inner gradients in isophote position
angle for the intermediate and power-law galaxies are plotted as a function
of the local logarithmic surface brightness slope, $\gamma',$ at the
radius dividing the inner and outer gradients.  The wedge symbols
also encode the value of the inner gradients in the slope of their
left line segments and the outer gradients in the right segments.
Dotted symbols are those galaxies that may be strongly affected by dust.
Radii of the break or ``hinge'' in the PA gradients cluster near
$\gamma'=1,$ larger than for core galaxies (Figure \ref{fig:bend}).
PA twists are also smaller than for core galaxies, but the text
shows that the relative amounts of light involved in the twist
phenomenon is the same in both types of galaxy.}
\label{fig:bend_pow}
\end{figure}
\begin{figure}
\plotone{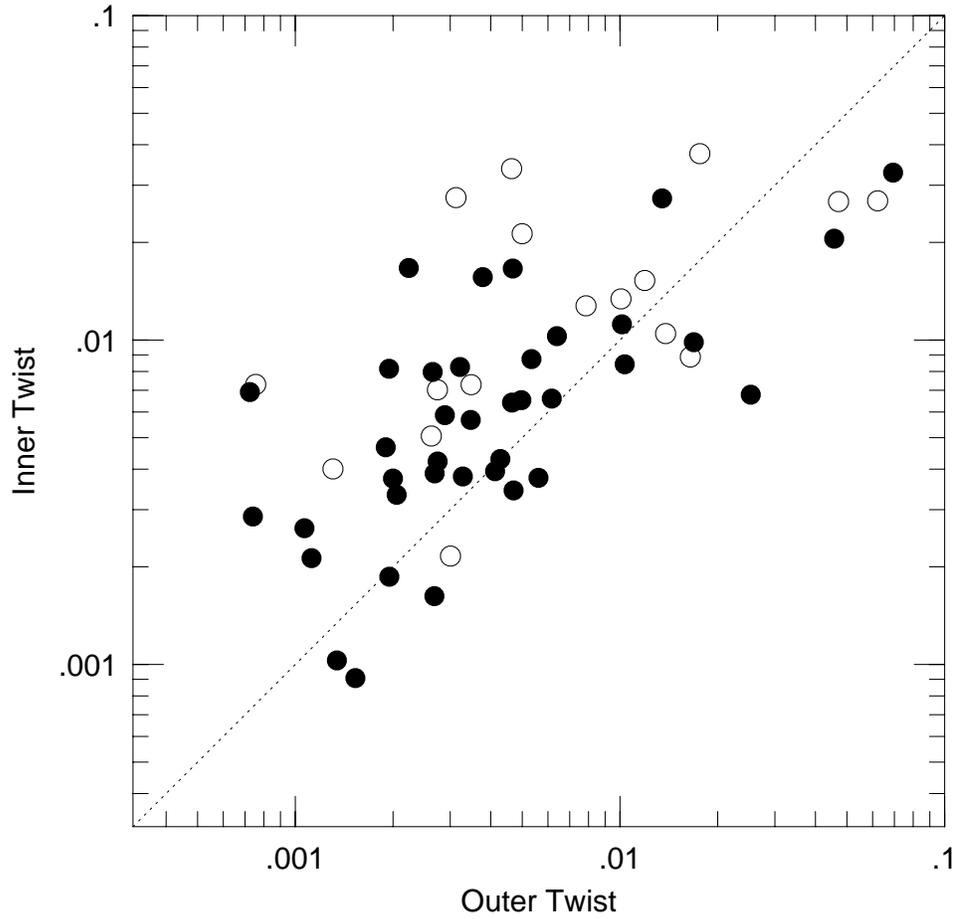}
\caption{Position angle twists defined using equation (\ref{eqn:twist})
for inner radii are plotted versus the twists for the outer radii.
Solid symbols are core galaxies; open symbols are power-law or intermediate
galaxies.  Twists are larger in inner regions, but the intrinsic
strength of twists in core and power-law galaxies is similar.}
\label{fig:twist}
\end{figure}
\begin{figure}
\plotone{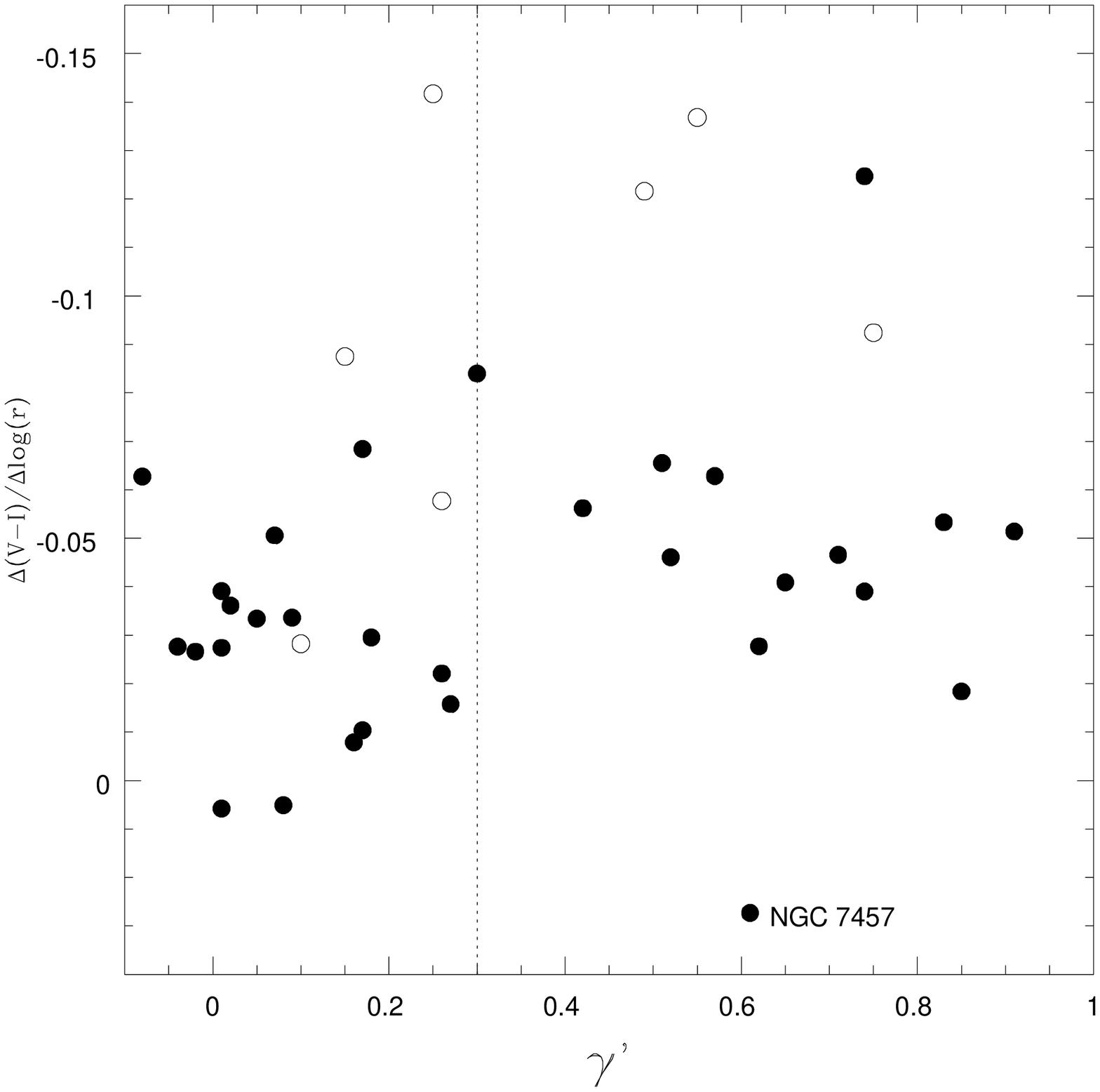}
\caption{Slopes of the inner $V-I$ color gradients as a function
of $\gamma';$ the vertical line separates core from power-law
and intermediate galaxies.  Open symbols are those galaxies
that may be strongly affected by dust.  After eliminating dusty galaxies,
power-law galaxies are seen to have gradients only slightly steeper
than core galaxies.}
\label{fig:grad_gam}
\end{figure}
\begin{figure}
\plotone{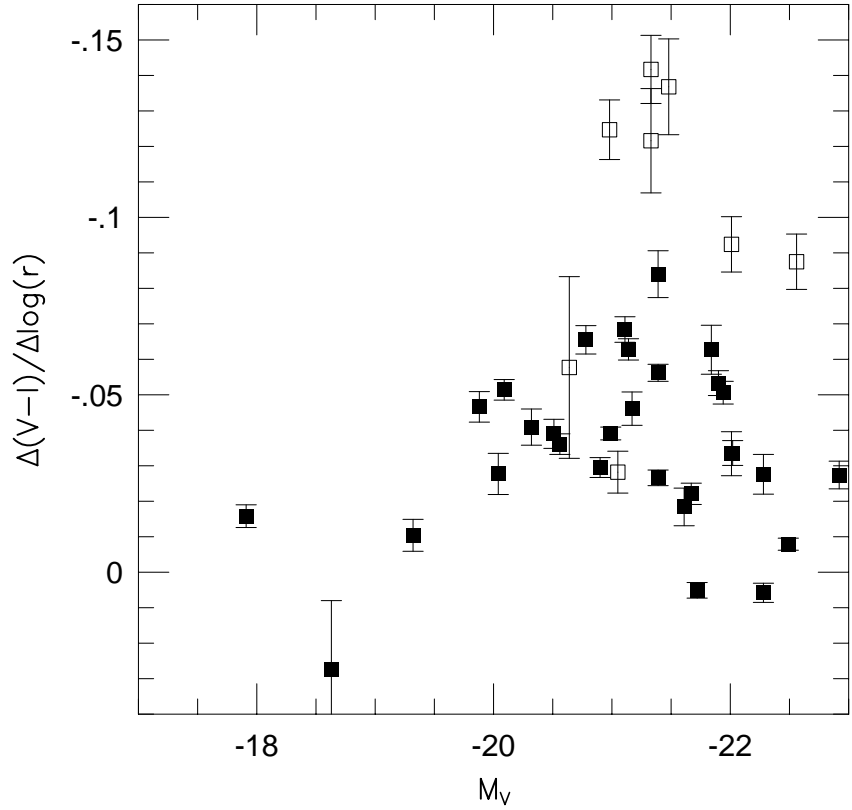}
\caption{Slopes of the inner $V-I$ color gradients as a function
galaxy luminosity.  Open symbols are those galaxies
that may be strongly affected by dust.  After these are eliminated,
there is little remaining trend in color gradient versus absolute magnitude.}
\label{fig:grad_mag}
\end{figure}
\begin{figure}
\plotone{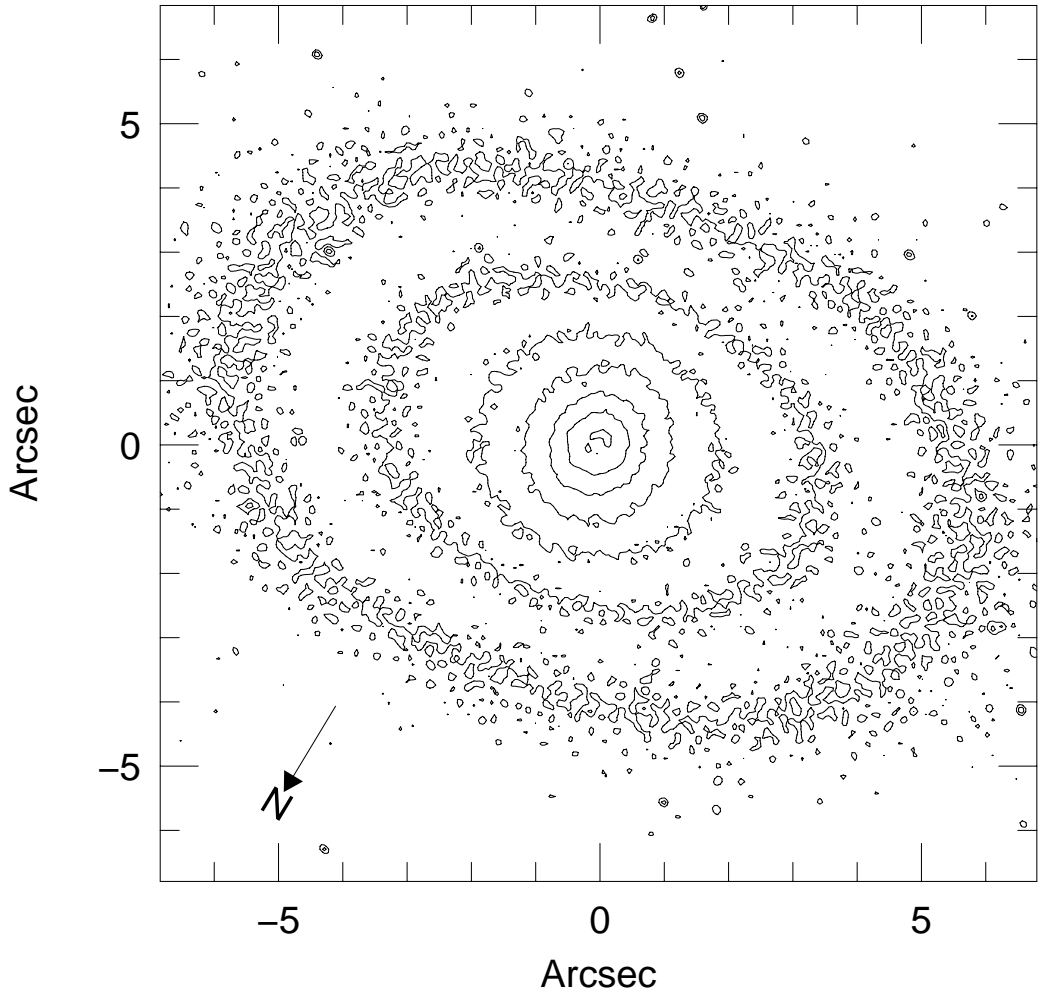}
\caption{Contour plot ($V$ band) of the center of the central-minimum
galaxy NGC 4073.  The contours are spaced in 0.5 mag increments; the
innermost contour is at $\mu_V=16.43$ and has been selected to highlight
asymmetries in the ring of light surrounding the central minimum.}
\label{fig:n4073_con}
\end{figure}
\begin{figure}
\plotone{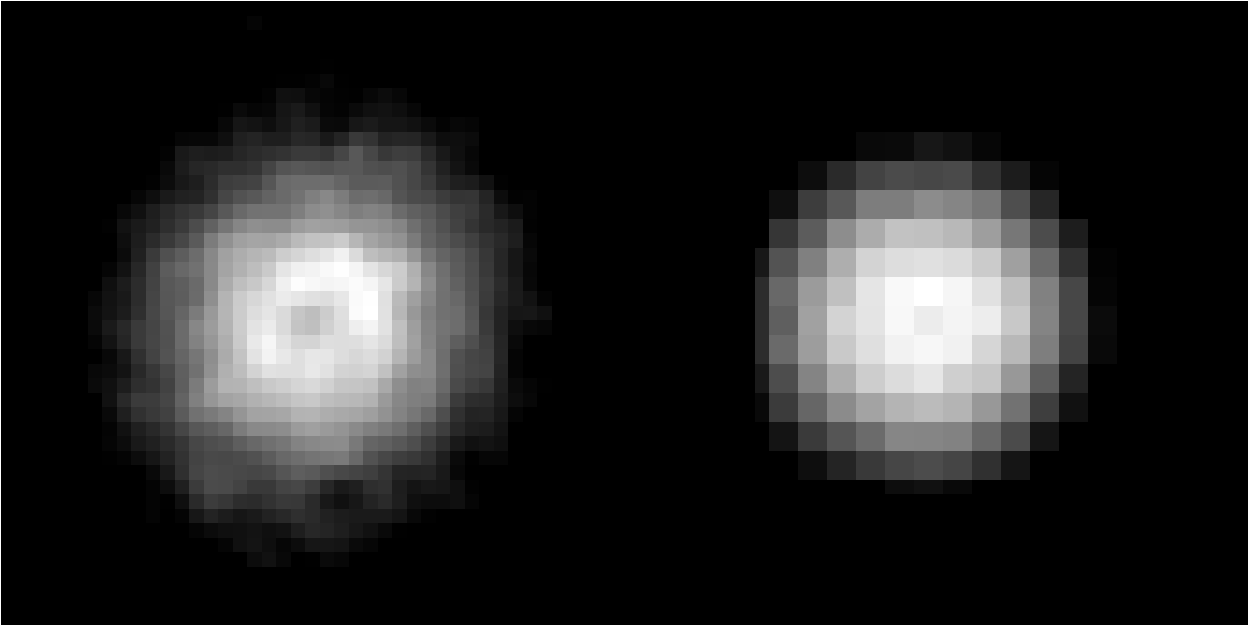}
\caption{Central images of the central-minimum galaxy NGC 4073.  The image
on the left is the central $2''\times2''$ region of the deconvolved WFPC2 F555W
image.  A hard contrast stretch has been used to bring out the
central minimum.  The image on the right is the NICMOS-2 $H$ band image
(from program GTO 7820)
as observed.  The NICMOS-2 pixels have been doubled in size to approximate
the WFPC2 scale.  The orientation of both images is arbitrary.  The presence
of the minimum in $H$ band suggests that it is not due to dust.}
\label{fig:n4073_im}
\end{figure}
\begin{figure}
\plotone{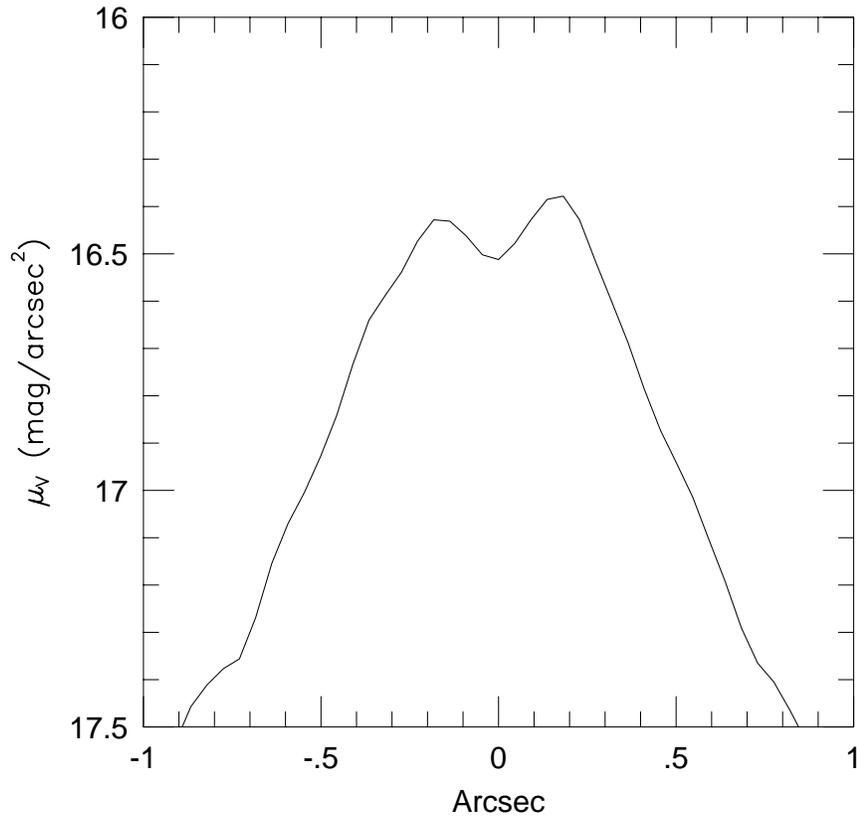}
\caption{This figure shows a 3-pixel wide cut taken through
the center of the central-minimum galaxy  NGC 4073 ($V$ band) at position angle $166^\circ.$}
\label{fig:n4073_cut}
\end{figure}
\begin{figure}
\plotone{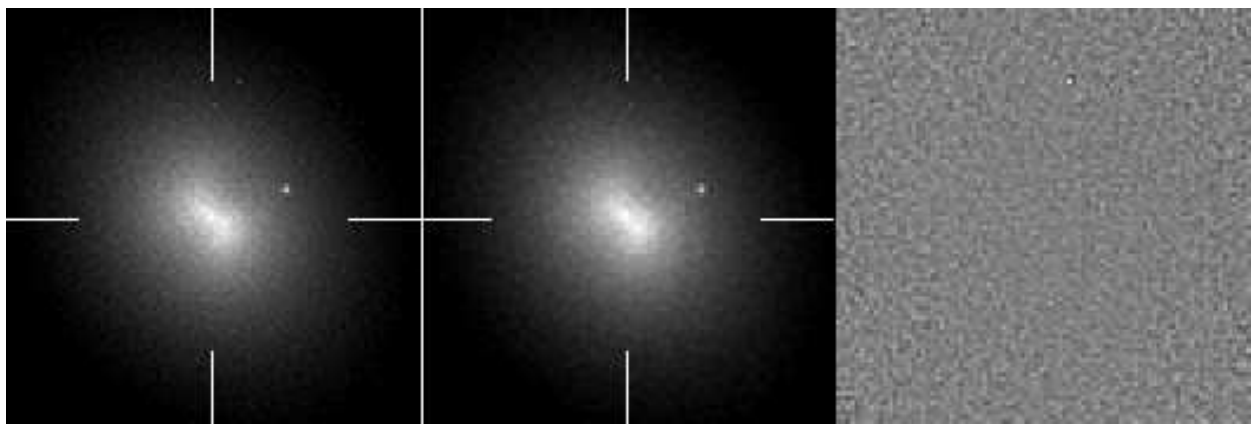}
\caption{The central $4''\times4''$ portions of the deconvolved
$V$ and $I$ band images of the central-minimum galaxy NGC 4382 are shown.
Tickmarks indicate the galaxy center.  The panel on the right is a ratio of
the $V$ to $I$ image, showing that there is no central dust absorption.
See Figure \ref{fig:n4382_con} for orientation.}
\label{fig:n4382_im}
\end{figure}
\begin{figure}
\plotone{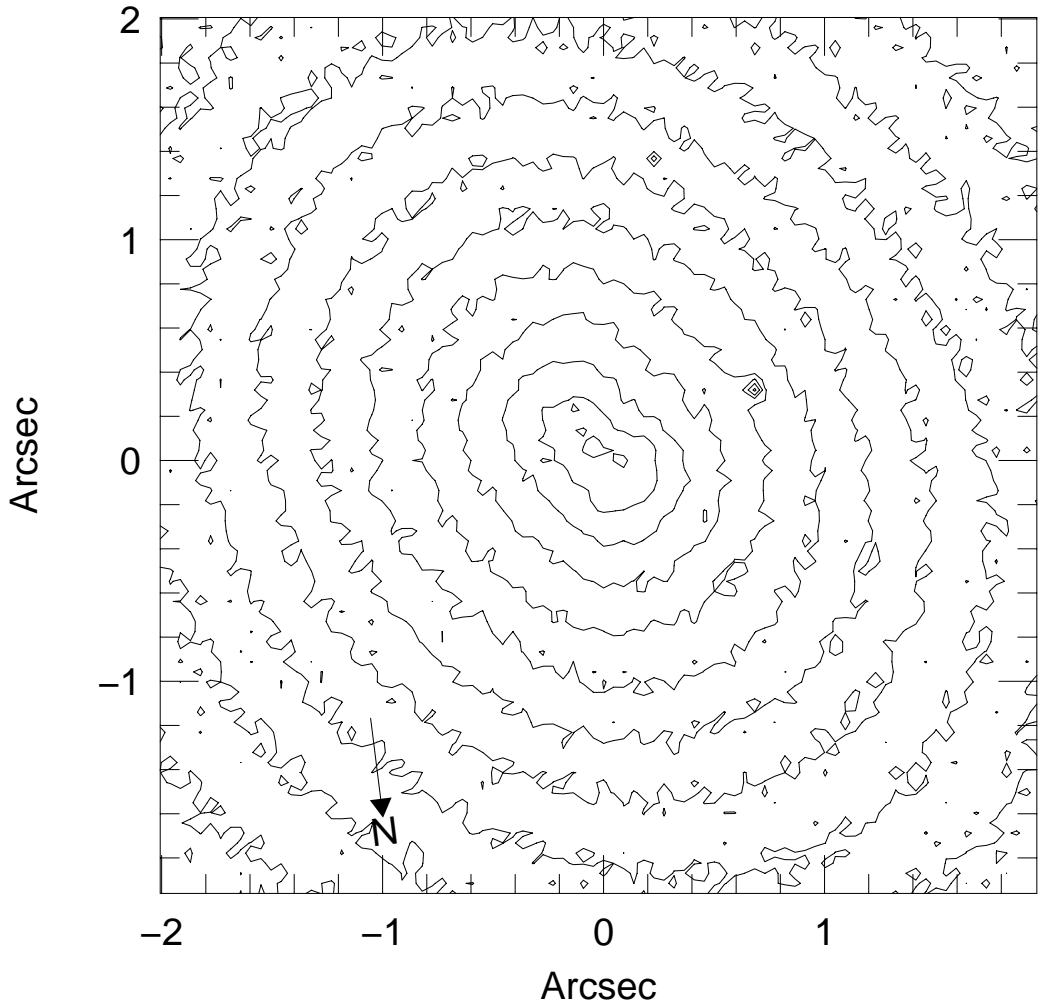}
\caption{Contour plot of the center of the deconvolved $V$ band image of the
central-minimum galaxy NGC 4382. Contours are in steps of 0.2 mag in
surface brightness.  The innermost contour is at $\mu_V=14.84$ and is
selected to highlight the peaks of the double nucleus.}
\label{fig:n4382_con}
\end{figure}
\begin{figure}
\plotone{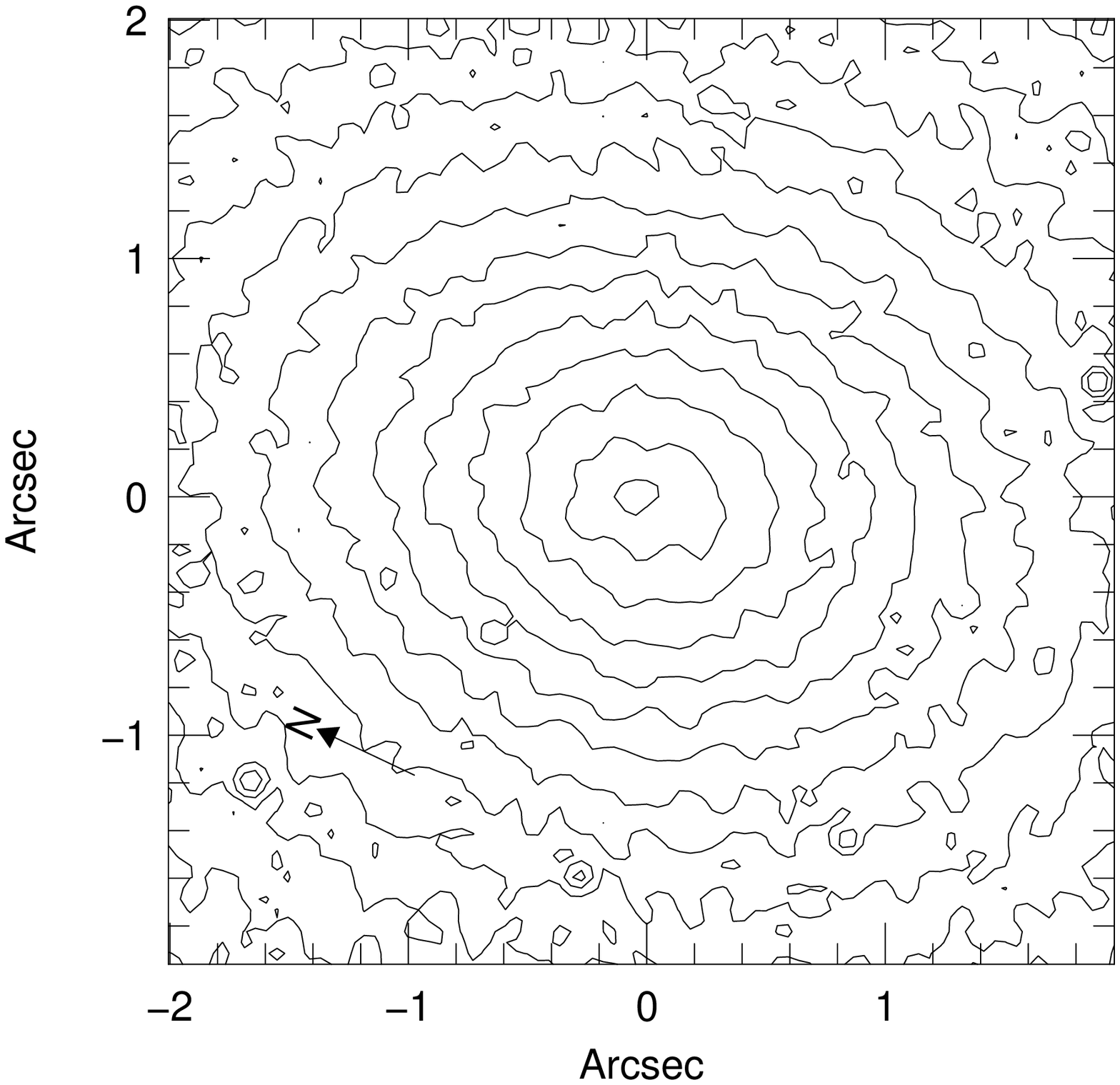}
\caption{Contour plot of the center of the deconvolved $V$ band
image of the offset-center galaxy NGC 507. Contours are in steps of 0.2 mag
in surface brightness.  The innermost contour is at $\mu_V=16.24$ and is
selected to highlight the central asymmetry of the brightness distribution.
It may be that NGC 507 is like M31 or NGC 4382 seen at lower resolution.}
\label{fig:n0507_con}
\end{figure}
\begin{figure}
\plotone{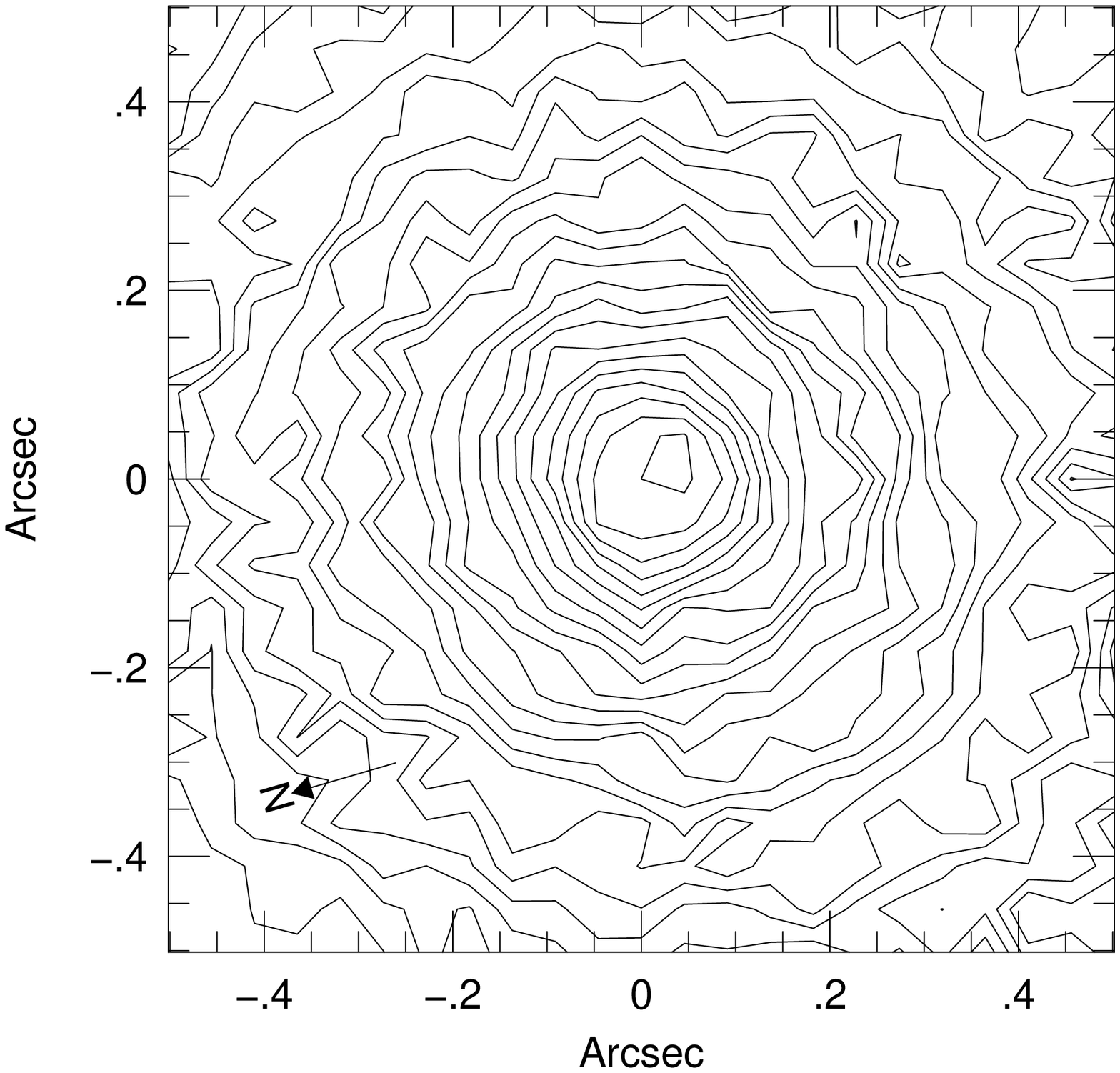}
\caption{Contour plot of the center of the deconvolved $V$ band
image of the offset-center galaxy NGC 1374. Contours are in steps of 0.1 mag
in surface brightness.  The innermost contour is at $\mu_R=13.73$ and is
selected to highlight the central asymmetry of the brightness distribution.
It may be that NGC 1374 is like M31 or NGC 4382 seen at lower resolution.}
\label{fig:n1374_con}
\end{figure}
\clearpage
\begin{figure}
\plotone{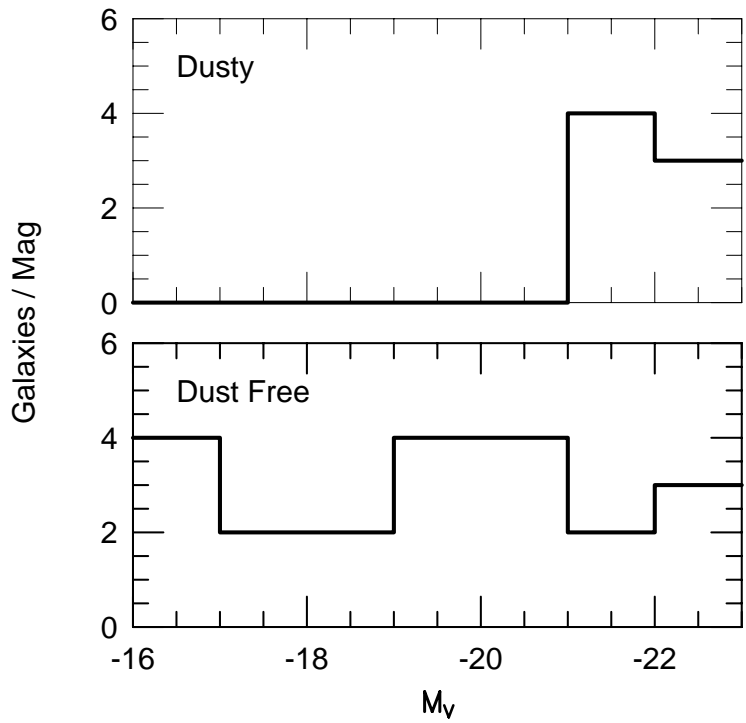}
\caption{Occurrence of dust clouds
in Virgo cluster elliptical galaxies is shown as a function of luminosity.
Histogram bins are one magnitude wide.  All galaxies fainter than $M_V=-21$
in Virgo are dust free.}
\label{fig:dust_mag}
\end{figure}
\begin{figure}
\plotone{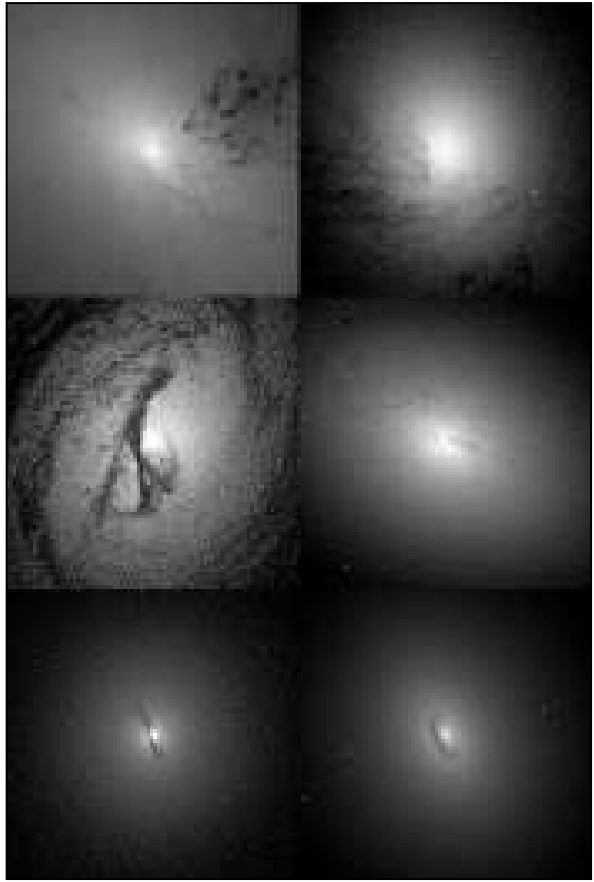}
\caption{Six galaxies have been selected to illustrate a hypothetical
dust-settling sequence.  NGC 1316 is the galaxy in the
top left panel and has the most poorly organized dust.  Top
right is NGC 4278, in which the dust appears to be streaming
towards the nucleus.  Center panels are NGC 3607 (left) and
IC 1459, in which a central dust ring is starting
to be organized.  Bottom left is NGC 2434, which has a
slightly warped ring.  NGC 4494, having a symmetric ring,
is the final galaxy in the sequence. Each panel is $8''$ on a side;
deconvolved images have been used.}
\label{fig:sequence}
\end{figure}
\begin{figure}
\plotone{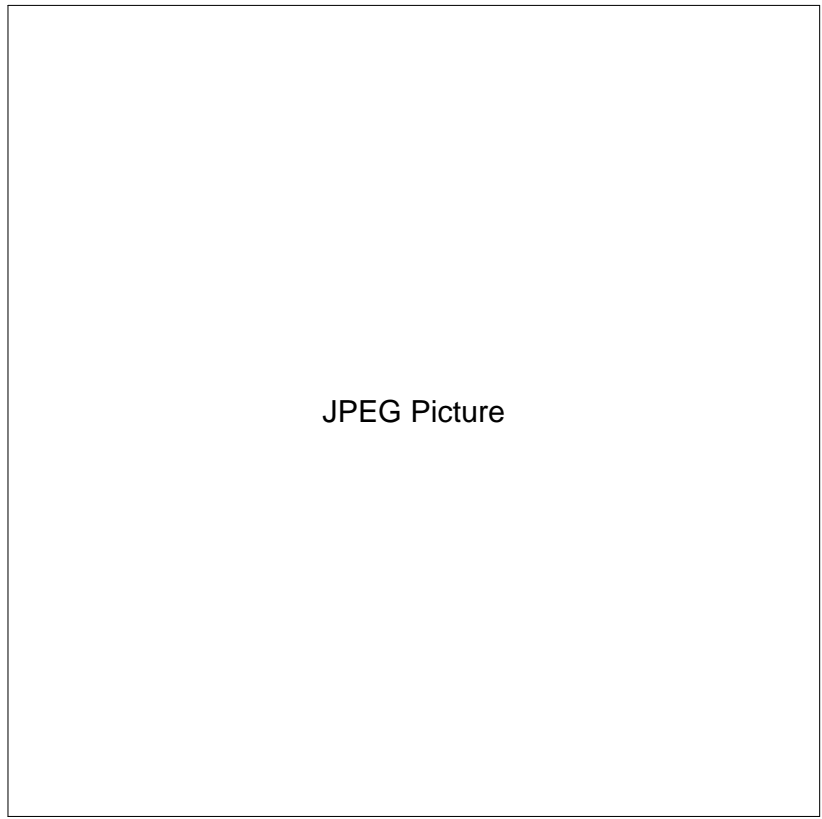}
\caption{The deconvolved V-band image of NGC 3607. A
logarithmic intensity stretch has been used.  The image
is 512 pixels or $23\asec3$ on a side.  The outer dust pattern is
remarkably symmetric and tightly wrapped, signifying a dust disk with
a well-defined plane and great age.  By contrast, the inner dust seems
to be settling to a plane that is {\it highly inclined} to the outer
plane and gives the appearance of being dynamically much younger.
Nevertheless, the two structures appear to join smoothly.}
\label{fig:n3607}
\end{figure}
\end{document}